\documentclass[twocolumn]{aastex701}

\newcommand{\kms}{\ensuremath{\mathrm{km\,s^{-1}}}}

\usepackage{CJKutf8}

\begin{document}
\begin{CJK*}{UTF8}{gbsn}

\title{No Evidence of Cold Trapping in the Dayside Atmosphere of WASP-33~b}

\author[orcid=0009-0005-5145-5165,sname=Choi]{Yeon-Ho Choi}
\affiliation{Korea Astronomy and Space Science Institute, 776 Daedeok-daero, Yuseong-gu, Daejeon 34055, Republic of Korea}
\email[show]{yhc@kasi.re.kr}  

\author[0000-0001-6876-8559]{Changwoo Kye}
\affiliation{Department of Physics and Astronomy, Seoul National University, Seoul 08826, Republic of Korea}
\email{fakeemail1@google.com}  

\author[0009-0005-4890-3326]{Krishna Kanumalla}
\affiliation{School of Earth and Space Exploration, Arizona State University, Tempe, AZ 85287, USA}
\email{skanuma1@asu.edu}  

\author[0000-0002-8573-805X]{Stefan Pelletier}
\affiliation{Observatoire astronomique de l'Université de Genève, 51 chemin Pegasi 1290 Versoix, Switzerland}
\affiliation{Institut Trottier de recherche sur les exoplan\`etes, D\'epartement de Physique, Universit\'e de Montr\'eal, Montr\'eal, Qu\'ebec, Canada}
\email{fakeemail1@google.com}  

\author[0000-0002-9946-5259]{Peter C. B. Smith}
\affiliation{School of Earth and Space Exploration, Arizona State University, Tempe, AZ 85287, USA}
\affiliation{Max-Planck Institut f\"ur Astronomie, Heidelberg 69117, Germany}
\email{fakeemail1@google.com}  

\author[0000-0002-2338-476X]{Michael R. Line}
\affiliation{School of Earth and Space Exploration, Arizona State University, Tempe, AZ 85287, USA}
\email{fakeemail1@google.com}  

\author[0000-0002-9142-6378]{Jorge A. Sanchez}
\affiliation{School of Earth and Space Exploration, Arizona State University, Tempe, AZ 85287, USA}
\email{fakeemail1@google.com}  

\author[0000-0003-0894-7824]{Jae-Joon Lee}
\affiliation{Korea Astronomy and Space Science Institute, 776 Daedeok-daero, Yuseong-gu, Daejeon 34055, Republic of Korea}
\email{fakeemail1@google.com}  

\author[0000-0002-2783-0755]{Byeong-Cheol Lee}
\affiliation{Korea Astronomy and Space Science Institute, 776 Daedeok-daero, Yuseong-gu, Daejeon 34055, Republic of Korea}
\affiliation{Department of Astronomy and Space Science, University of Science and Technology, 217 Gajeong-ro, Yuseong-gu, Daejeon 34113, Republic of Korea}
\email{fakeemail1@google.com}

\author[0000-0003-4698-6285]{Stevanus K. Nugroho}
\affiliation{Department of Earth and Planetary Sciences, School of Science, Institute of Science Tokyo, 2-12-1 Ookayama, Meguro, Tokyo 152-8551, Japan}
\affiliation{Astrobiology Center, NINS, 2-21-1 Osawa, Mitaka, Tokyo 181-8588, Japan}
\email{nugroho@eps.sci.isct.ac.jp}

\author[orcid=0000-0001-8877-0242,sname='Hayashi']{Yuya Hayashi}
\affiliation{Department of Multi-Disciplinary Sciences, Graduate School of Arts and Sciences, The University of Tokyo, 3-8-1 Komaba, Meguro, Tokyo 153-8902, Japan}
\email{yuya-hayashi@g.ecc.u-tokyo.ac.jp}

\author[0000-0003-0815-8366]{Swaetha Ramkumar}
\affiliation{School of Earth and Space Exploration, Arizona State University, Tempe, AZ 85287, USA}
\email{fakeemail1@google.com}

\author[0000-0002-2513-4465]{Vatsal Panwar}
\affiliation{School of Physics \& Astronomy, University of Birmingham, Edgbaston, Birmingham B15 2TT, UK}
\email{fakeemail1@google.com}

\author[0000-0003-3860-6297]{Huan-Yu Teng}
\affiliation{National Astronomical Observatories, Chinese Academy of Sciences, Beijing 100101, PR China}
\email{fakeemail1@google.com}

\author[0000-0001-9585-9034]{Fei Yan}
\affiliation{Department of Astronomy, University of Science and Technology of China, Hefei 230026, PR China}
\affiliation{Deep Space Exploration Laboratory, Hefei, Anhui 230026, China}
\email{fakeemail1@google.com}

\author[0000-0002-4361-8885]{Ji Wang (王吉)}
\affiliation{Department of Astronomy, The Ohio State University, 100 West 18th Avenue, Columbus, OH 43210, USA}
\email{fakeemail1@google.com}

\author[0000-0001-8969-0009]{Jae-Rim Koo}
\affiliation{Korea Astronomy and Space Science Institute, 776 Daedeok-daero, Yuseong-gu, Daejeon 34055, Republic of Korea}
\email{fakeemail1@google.com}

\author[0000-0002-4477-1208]{Jangho Bae}
\affiliation{Department of Physics and Astronomy, Seoul National University, Seoul 08826, Republic of Korea}
\affiliation{SNU Astronomy Research Center, Astronomy Program, Department of Physics \& Astronomy, Seoul National University, 1 Gwanak-ro, Gwanak-gu, Seoul 08826, Republic of Korea}
\email{fakeemail1@google.com}

\author[0000-0001-9773-3080]{Chan Park}
\affiliation{Korea Astronomy and Space Science Institute, 776 Daedeok-daero, Yuseong-gu, Daejeon 34055, Republic of Korea}
\email{fakeemail1@google.com}  

\author[0000-0002-0418-5335]{Heeyoung Oh}
\affiliation{Korea Astronomy and Space Science Institute, 776 Daedeok-daero, Yuseong-gu, Daejeon 34055, Republic of Korea}
\email{fakeemail1@google.com}  

\author{Sanghyuk Kim}
\affiliation{Korea Astronomy and Space Science Institute, 776 Daedeok-daero, Yuseong-gu, Daejeon 34055, Republic of Korea}
\email{fakeemail1@google.com}  

\author[0000-0002-4445-7366]{Ueejeong Jeong}
\affiliation{Korea Astronomy and Space Science Institute, 776 Daedeok-daero, Yuseong-gu, Daejeon 34055, Republic of Korea}
\affiliation{University of Texas at Austin, 2515 Speedway, Stop C1400, Austin, Texas 78712-1205, USA}
\email{fakeemail1@google.com}  

\author[0000-0003-0871-3665]{Hye-In Lee}
\affiliation{Korea Astronomy and Space Science Institute, 776 Daedeok-daero, Yuseong-gu, Daejeon 34055, Republic of Korea}
\email{fakeemail1@google.com}  

\author[0000-0001-8012-5871]{Woojin Park}
\affiliation{Korea Astronomy and Space Science Institute, 776 Daedeok-daero, Yuseong-gu, Daejeon 34055, Republic of Korea}
\email{fakeemail1@google.com}  

\author{Youngsam Yu}
\affiliation{Korea Astronomy and Space Science Institute, 776 Daedeok-daero, Yuseong-gu, Daejeon 34055, Republic of Korea}
\email{fakeemail1@google.com}  

\author{Yunjong Kim}
\affiliation{Korea Astronomy and Space Science Institute, 776 Daedeok-daero, Yuseong-gu, Daejeon 34055, Republic of Korea}
\email{fakeemail1@google.com}  

\author{Moo-Young Chun}
\affiliation{Korea Astronomy and Space Science Institute, 776 Daedeok-daero, Yuseong-gu, Daejeon 34055, Republic of Korea}
\email{fakeemail1@google.com}  

\author{Jae Sok Oh}
\affiliation{Korea Astronomy and Space Science Institute, 776 Daedeok-daero, Yuseong-gu, Daejeon 34055, Republic of Korea}
\email{fakeemail1@google.com}  

\author{Sungho Lee}
\affiliation{Korea Astronomy and Space Science Institute, 776 Daedeok-daero, Yuseong-gu, Daejeon 34055, Republic of Korea}
\email{fakeemail1@google.com}  

\author{Jeong-Gyun Jang}
\affiliation{Korea Astronomy and Space Science Institute, 776 Daedeok-daero, Yuseong-gu, Daejeon 34055, Republic of Korea}
\email{fakeemail1@google.com}  

\author{Bi-Ho Jang}
\affiliation{Korea Astronomy and Space Science Institute, 776 Daedeok-daero, Yuseong-gu, Daejeon 34055, Republic of Korea}
\email{fakeemail1@google.com}  

\author{Hyeon Cheol Seong}
\affiliation{Korea Astronomy and Space Science Institute, 776 Daedeok-daero, Yuseong-gu, Daejeon 34055, Republic of Korea}
\email{fakeemail1@google.com}

\author[0000-0001-9263-3275]{Hyun-Jeong Kim}
\affiliation{Korea Astronomy and Space Science Institute, 776 Daedeok-daero, Yuseong-gu, Daejeon 34055, Republic of Korea}
\email{fakeemail1@google.com}  

\author{Cynthia B. Brooks}
\affiliation{University of Texas at Austin, 2515 Speedway, Stop C1400, Austin, Texas 78712-1205, USA}
\email{fakeemail1@google.com}  

\author[0000-0001-7875-6391]{Gregory N. Mace}
\affiliation{University of Texas at Austin, 2515 Speedway, Stop C1400, Austin, Texas 78712-1205, USA}
\email{fakeemail1@google.com}  

\author{Hanshin Lee}
\affiliation{University of Texas at Austin, 2515 Speedway, Stop C1400, Austin, Texas 78712-1205, USA}
\email{fakeemail1@google.com}  

\author{John M. Good}
\affiliation{University of Texas at Austin, 2515 Speedway, Stop C1400, Austin, Texas 78712-1205, USA}
\email{fakeemail1@google.com}  

\author[0000-0003-3577-3540]{Daniel T. Jaffe}
\affiliation{University of Texas at Austin, 2515 Speedway, Stop C1400, Austin, Texas 78712-1205, USA}
\email{fakeemail1@google.com}  

\author{Kang-Min Kim}
\affiliation{Korea Astronomy and Space Science Institute, 776 Daedeok-daero, Yuseong-gu, Daejeon 34055, Republic of Korea}
\email{fakeemail1@google.com}  

\author{In-Soo Yuk}
\affiliation{Korea Astronomy and Space Science Institute, 776 Daedeok-daero, Yuseong-gu, Daejeon 34055, Republic of Korea}
\email{fakeemail1@google.com}  

\author[0000-0002-2013-1273]{Narae Hwang}
\affiliation{Korea Astronomy and Space Science Institute, 776 Daedeok-daero, Yuseong-gu, Daejeon 34055, Republic of Korea}
\email{fakeemail1@google.com}  

\author[0000-0002-6982-7722]{Byeong-Gon Park}
\affiliation{Korea Astronomy and Space Science Institute, 776 Daedeok-daero, Yuseong-gu, Daejeon 34055, Republic of Korea}
\email{fakeemail1@google.com}  

\author[0000-0003-4770-688X]{Hwihyun Kim}
\affiliation{International Gemini Observatory/NSF NOIRLab, 950 N. Cherry Ave., Tucson, AZ 85719, USA}
\email{fakeemail1@google.com}  

\author{Brian Chinn}
\affiliation{International Gemini Observatory/NSF NOIRLab, Casilla 603, La Serena, Chile}
\email{fakeemail1@google.com}  

\author{Francisco Ramos}
\affiliation{International Gemini Observatory/NSF NOIRLab, Casilla 603, La Serena, Chile}
\email{fakeemail1@google.com}  

\author{Pablo Prado}
\affiliation{International Gemini Observatory/NSF NOIRLab, Casilla 603, La Serena, Chile}
\email{fakeemail1@google.com}  

\author{Ruben Diaz}
\affiliation{International Gemini Observatory/NSF NOIRLab, Casilla 603, La Serena, Chile}
\email{fakeemail1@google.com}  

\author{John White}
\affiliation{International Gemini Observatory/NSF NOIRLab, 670 N. A’ohoku Place, Hilo, Hawai’i, 96720, USA}
\email{fakeemail1@google.com}  

\author{Eduardo Tapia}
\affiliation{International Gemini Observatory/NSF NOIRLab, 670 N. A’ohoku Place, Hilo, Hawai’i, 96720, USA}
\email{fakeemail1@google.com}  

\author{Andres Olivares}
\affiliation{International Gemini Observatory/NSF NOIRLab, Casilla 603, La Serena, Chile}
\email{fakeemail1@google.com}  

\author{Valentina Oyarzun}
\affiliation{International Gemini Observatory/NSF NOIRLab, Casilla 603, La Serena, Chile}
\email{fakeemail1@google.com}  

\author{Emma Kurz}
\affiliation{International Gemini Observatory/NSF NOIRLab, 670 N. A’ohoku Place, Hilo, Hawai’i, 96720, USA}
\email{fakeemail1@google.com}  

\author{Hawi Stecher}
\affiliation{International Gemini Observatory/NSF NOIRLab, 670 N. A’ohoku Place, Hilo, Hawai’i, 96720, USA}
\email{fakeemail1@google.com}  

\author{Carlos Quiroz}
\affiliation{International Gemini Observatory/NSF NOIRLab, Casilla 603, La Serena, Chile}
\email{fakeemail1@google.com}  

\author{Ignacio Arriagada}
\affiliation{International Gemini Observatory/NSF NOIRLab, Casilla 603, La Serena, Chile}
\email{fakeemail1@google.com}  

\author{Thomas L. Hayward}
\affiliation{International Gemini Observatory/NSF NOIRLab, Casilla 603, La Serena, Chile}
\email{fakeemail1@google.com}  

\author[0000-0002-2536-1633]{Hyewon Suh}
\affiliation{International Gemini Observatory/NSF NOIRLab, 670 N. A’ohoku Place, Hilo, Hawai’i, 96720, USA}
\email{fakeemail1@google.com}  

\author{Jen Miller}
\affiliation{International Gemini Observatory/NSF NOIRLab, 670 N. A’ohoku Place, Hilo, Hawai’i, 96720, USA}
\email{fakeemail1@google.com}  

\author{Siyi Xu}
\affiliation{International Gemini Observatory/NSF NOIRLab, 670 N. A’ohoku Place, Hilo, Hawai’i, 96720, USA}
\email{fakeemail1@google.com}  

\author[0000-0002-6822-2254]{Emanuele Paolo Farina}
\affiliation{International Gemini Observatory/NSF NOIRLab, 670 N. A’ohoku Place, Hilo, Hawai’i, 96720, USA}
\email{fakeemail1@google.com}  

\author{Charlie Figura}
\affiliation{International Gemini Observatory/NSF NOIRLab, 670 N. A’ohoku Place, Hilo, Hawai’i, 96720, USA}
\email{fakeemail1@google.com}  

\author[0000-0003-4603-556X]{Teo Mocnik}
\affiliation{International Gemini Observatory/NSF NOIRLab, 670 N. A’ohoku Place, Hilo, Hawai’i, 96720, USA}
\email{fakeemail1@google.com}  

\author[0000-0003-4236-6927]{Zachary Hartman}
\affiliation{International Gemini Observatory/NSF NOIRLab, 670 N. A’ohoku Place, Hilo, Hawai’i, 96720, USA}
\affiliation{NASA Ames Research Center, Moffett Field, CA 94035, USA}
\email{fakeemail1@google.com}  

\author[0000-0002-6529-202X]{Mark Rawlings}
\affiliation{National Radio Astronomy Observatory, 520 Edgemont Road, Charlottesville, VA, 22903, USA}
\affiliation{International Gemini Observatory/NSF NOIRLab, 670 N. A’ohoku Place, Hilo, Hawai’i, 96720, USA}
\email{fakeemail1@google.com}  

\author{Andrew Stephens}
\affiliation{International Gemini Observatory/NSF NOIRLab, 670 N. A’ohoku Place, Hilo, Hawai’i, 96720, USA}
\email{fakeemail1@google.com}  

\author[0000-0002-5665-376X]{Bryan Miller}
\affiliation{International Gemini Observatory/NSF NOIRLab, Casilla 603, La Serena, Chile}
\email{fakeemail1@google.com}  

\author[0000-0002-6633-7891]{Kathleen Labrie}
\affiliation{International Gemini Observatory/NSF NOIRLab, 670 N. A’ohoku Place, Hilo, Hawai’i, 96720, USA}
\email{fakeemail1@google.com}  

\author{Paul Hirst}
\affiliation{International Gemini Observatory/NSF NOIRLab, 670 N. A’ohoku Place, Hilo, Hawai’i, 96720, USA}
\email{fakeemail1@google.com}

\begin{abstract}

Ultra-hot Jupiters allow simultaneous measurements of volatile molecules and refractory metals, providing constraints on atmospheric chemistry and refractory-element sequestration. We reanalyze the previously published Gemini-N/IGRINS-2 dayside emission spectra of WASP-33~b, extending the analysis to neutral refractory species. 
Using cross-correlation (CC) and likelihood-based (LL) detection methods, we recover the strongest species detection signal-to-noise ratios (S/N) from CO ($\rm CC~S/N=5.29,\ LL~S/N=5.62$). \ion{Mg}{1} ($\rm CC~S/N=3.88,\ LL~S/N=4.73$) and OH ($\rm CC~S/N=3.88,\ LL~S/N=3.89$) are recovered at lower significance in both metrics. H$_2$O remains marginal, appearing only in the cross-correlation analysis ($\rm CC~S/N=3.24$), while \ion{Ti}{1} ($\rm LL~S/N=3.70$) and \ion{Ca}{1} ($\rm LL~S/N=2.98$) show tentative signals. We do not detect \ion{Fe}{1}, \ion{Si}{1}, and \ion{V}{1} in this dataset.
A chemical-equilibrium retrieval with a flexible Bezier-spline pressure--temperature profile yields ${\rm C/O}=0.65^{+0.09}_{-0.10}$ and $[{\rm M/H}]=-0.48^{+0.24}_{-0.18}$, and a refractory-to-volatile ratio of $[R/V]_{\rm p}=-0.15^{+0.24}_{-0.23}$. 
The corresponding stellar-normalized value is $[R/V]_{\rm p/ \star}=0.37^{+0.38}_{-0.36}$, but this value is sensitive to the stellar oxygen abundance.
We find that the atmospheric Ti/Mg and Ca/Mg ratios of WASP-33~b are consistent with their respective stellar values within 68\% confidence. This is in contrast to other ultra-hot Jupiters with lower equilibrium temperatures that often show depleted Ti abundances compared to other metals, indicative of cold-trapping.  This would suggest that even the nightside of WASP-33~b is too hot for significant condensation and sequestration to occur.
Comparison with disk-chemistry models remains exploratory because the carbon-to-refractory posterior overlaps several formation scenarios. 


\end{abstract}



\section{Introduction} 

Atmospheric elemental abundances provide a means of tracing the formation and migration histories of hot Jupiters. The carbon-to-oxygen ratio (C/O) and overall metallicity depend on where a planet accreted gas and solids relative to major condensation fronts such as the H$_2$O, CO$_2$, and CO snow lines \citep{Oberg2011, Madhusudhan2012}. However, C/O alone is degenerate because similar values can arise from different combinations of gas accretion, solid accretion, and elemental sequestration \citep{Lothringer2021, Turrini2021, Chachan2023}. Refractory elements provide an additional constraint by tracing the rocky material incorporated into the planetary atmosphere.

Ultra-hot Jupiters (UHJs; T$_{\rm eq}$ $\geq$2000 K) are especially useful for this test. Their high atmospheric temperatures keep refractory species in the gas phase, enabling simultaneous measurements of volatile molecules and refractory metals \citep{Lothringer2021}. These conditions also reduce the likelihood of high-altitude condensate clouds and therefore lead to relatively clear spectra for transmission and emission observations. High-resolution spectroscopy of UHJs has simultaneously detected refractory metals (e.g., Fe, Mg, Ca) and volatile molecules such as CO, H$_{2}$O, and OH \citep{Ramkumar2023, Smith2024b, Bazinet2025, Yang2025, Sanchez2026, Kanumalla2026}. Abundance constraints obtained from such detections provide important constraints on both atmospheric chemistry and possible formation scenarios.

Refractory species also provide a test of atmospheric cold trapping. 
In a tidally locked hot Jupiter, the dayside can be hot enough to keep metals in the gas phase, while the nightside or deeper atmosphere may still be cool enough for condensates to form and rain out \citep{Parmentier2013}. If condensates settle before being transported back to the dayside, the observable atmosphere can become depleted in the corresponding refractory element. Ti is an especially sensitive diagnostic because it is easier to condense than Fe, Mg, and Si: Ti condenses at \(T_{\rm cond}\sim1600\)~K, compared with \(T_{\rm cond}\sim1350\)--1400~K for Fe, Mg, and Si. Ti can be removed through Ti oxides and CaTiO$_3$ condensates \citep{Lodders2002, Lodders2003}.
Observationally, the UHJ WASP-76~b shows a sharp depletion of elements with \(T_{\rm cond}>1550\)~K, interpreted as nightside cold trapping \citep{Pelletier2023}. The slightly hotter UHJ WASP-121~b also shows evidence for Ti depletion from both low- and high-resolution spectroscopy \citep{Hoeijmakers2024, Smith2024b, Prinoth2025, Pelletier2026a}. Meanwhile, hotter UHJs such as WASP-189~b, HAT-P-70~b, and MASCARA-1~b show detectable Ti-bearing species, although partial Ti depletion may still occur depending on nightside temperature, heat redistribution, ionization, and vertical mixing \citep{Sanchez2026, Pelletier2026b, Guo2026, Sun2026, Kanumalla2026}. Overall, a comparative retrieval of six UHJs similarly found Ti-bearing species to be underabundant in most planets in the sample, potentially due to a combination of nightside rain-out and ionization, with Ti and TiO constrained only for the hottest target, WASP-189~b \citep{Gandhi2023}.

\begin{figure*}[ht!] \centering
\includegraphics[width=0.99\textwidth]{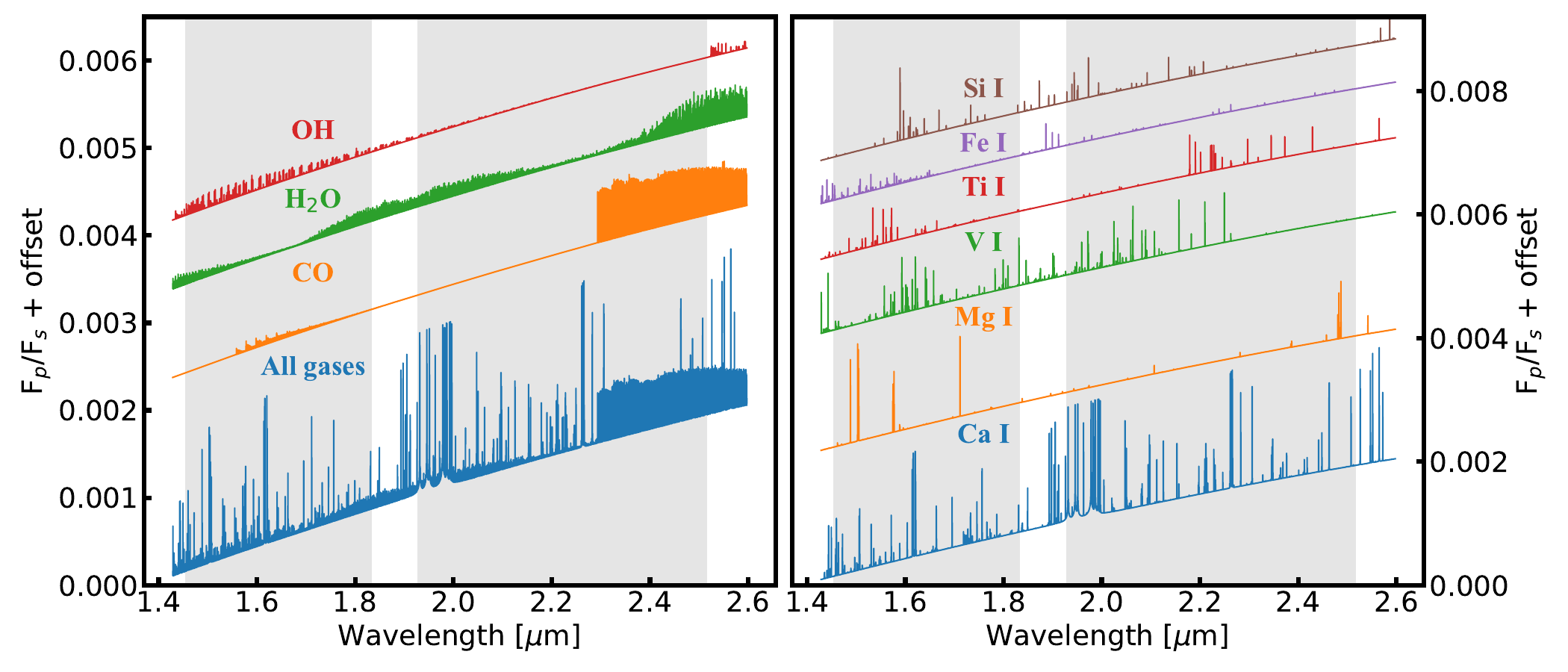}

\caption{
Thermal emission model spectra, shown as the planet-to-star flux ratio ${\rm F}_{p}/{\rm F}_{s}$, generated with CHIMERA using all included opacity sources. Shaded regions indicate the wavelength coverage of IGRINS-2. Individual opacity contributions from volatile molecules (H$_2$O, CO, OH) and refractory species (\ion{Fe}{1}, \ion{Ti}{1}, \ion{Mg}{1}, \ion{Ca}{1}, \ion{Si}{1}, \ion{V}{1}) are shown. These models form the basis for cross-correlation.
}
\label{fig:opacities}
\end{figure*}

WASP-33~b is a prototypical UHJ orbiting a $\delta$ Scuti A-type star every 1.22 days \citep{CollierCameron2010}. With a dayside temperature exceeding 2700~K \citep{Zhang2018, vonEssen2020}, it resides well within a regime where molecules can thermally dissociate and refractory elements are expected to be fully vaporized and part of the gas phase. Previous high-resolution spectroscopic observations have revealed the presence of CO, OH, and H$_{2}$O  \citep[e.g.,][]{Nugroho2021, Yan2022, vanSluijs2023, Finnerty2023} and neutral refractory gases (\ion{Fe}{1}, \ion{V}{1}, \ion{Ti}{1}, \ion{Si}{1}) \citep[e.g.,][]{Nugroho2020, Cont2021, Cont2022a, Cont2022b, Herman2022} in its dayside atmosphere. In addition to dayside observations, transit observations with various spectrographs have revealed the presence of ionized Ca in its atmosphere \citep{Yan2019, Yang2024a}. Some studies have also suggested evidence for nightside thermal emission of H$_{2}$O \citep{Yang2024b} and reported a detection of CO absorption \citep{Mraz2024}. Recently, \citet[][hereafter, C25]{Choi2025} used the upgraded Immersion GRating INfrared Spectrograph 2 \citep[IGRINS-2;][]{Lee2022, Oh2024} at Gemini-North to detect multiple volatile species in WASP-33~b's dayside emission spectra. This study also demonstrated the instrument's improved throughput and extended K-band coverage, both of which increase CO sensitivity.

In this work, we build on the analysis of C25 by using the same IGRINS-2 dataset to investigate both volatile molecules and refractory metals in WASP-33~b's atmosphere. We report a new detection of \ion{Mg}{1} and a tentative detection of \ion{Ca}{1} and \ion{Ti}{1} along with the previous detections of CO, H$_{2}$O, and OH. 
We use chemical-equilibrium retrievals to constrain the atmospheric elemental abundances and refractory-to-volatile ratio, examine relative refractory abundances as diagnostics of cold trapping, and compare the inferred composition with the disk-chemistry framework of \citet{Chachan2023}. We also place WASP-33~b in the emerging sample of misaligned UHJs with volatile and refractory abundance measurements.

\begin{table*}[t]
    \centering
    \caption{Stellar and planetary parameters and stellar abundances of the WASP-33 system.}
    \label{tab:system_parameters}
    \begin{tabular}{llclc}
    \hline\hline
        Parameter & Symbol & Value & Unit & Reference \\
        \hline
        \multicolumn{5}{l}{\textit{Stellar parameters}} \\
        Effective temperature & $T_{\rm eff}$ & $7430 \pm 100$ & K & 1 \\
        Mass & $M_\star$ & $1.495 \pm 0.031$ & $M_\odot$ & 1 \\
        Radius & $R_\star$ & $1.444 \pm 0.034$ & $R_\odot$ & 1 \\
        \hline
         & $[\mathrm{C/H}]$ & $-0.24 \pm 0.16$ & dex & 2 \\
         & $[\mathrm{O/H}]$ & $+0.58 \pm 0.23$ & dex & '' \\
        Stellar abundances & $[\mathrm{Fe/H}]$ & $+0.11 \pm 0.23$ & dex & '' \\
        (relative to solar value) & $[\mathrm{Mg/H}]$ & $-0.11 \pm 0.21$ & dex & '' \\
         & $[\mathrm{Si/H}]$ & $+0.11 \pm 0.20$ & dex & '' \\
         & $[\mathrm{Ca/H}]$ & $-0.20 \pm 0.17$ & dex & '' \\
         & $[\mathrm{Ti/H}]$ & $+0.08 \pm 0.16$ & dex & '' \\
        \hline
        \multicolumn{5}{l}{\textit{Planetary parameters}} \\
        Mass & $M_{\rm p}$ & $2.093 \pm 0.139$ & $M_{\rm Jup}$ & 1 \\
        Radius & $R_{\rm p}$ & $1.593 \pm 0.074$ & $R_{\rm Jup}$ & 1 \\
        Equilibrium temperature & $T_{\rm eq}$ & $2782 \pm 41$ & K & 1 \\
        \hline
        \multicolumn{5}{l}{\textit{Orbital and transit parameters}} \\
        Orbital period & $P$ & $1.219870 \pm 0.000001$ & day & 1 \\
        Semimajor axis & $a$ & $0.02390 \pm 0.00063$ & au & 1 \\
        Scaled semimajor axis & $a/R_\star$ & $3.571 \pm 0.010$ & --- & 1 \\
        Orbital inclination & $i$ & $86.63 \pm 0.03$ & deg & 1 \\
        Transit duration & $T_{14}$ & $2.854 \pm 0.012$ & hr & 1 \\
        \hline
        \multicolumn{5}{l}{\textit{Velocity reference values and model assumption}} \\
        Planetary RV semi-amplitude & $K_{\rm p}$ & $226.0$ & km\,s$^{-1}$ & 3 \\
        Systemic velocity & $V_{\rm sys}$ & $-9.2$ & km\,s$^{-1}$ & 4 \\
        Adopted equatorial rotation speed & $v_{\rm eq}$ & $6.64$ & km\,s$^{-1}$ & 1 \\
    \hline
    \end{tabular}
    \par\smallskip
    \begin{minipage}{\textwidth}
    \footnotesize
    \textit{Note.} The adopted equatorial rotation speed is based on
    $v_{\rm eq}=2\pi R_{\rm p}/P$, assuming synchronous rotation,
    with $R_{\rm p}$ and $P$ from Reference~1.
    \par
    \textit{References.} (1) \citet{Chakrabarty2019};
    (2) \citet{Saffe2021}; (3) \citet{Choi2025};
    (4) \citet{Gontcharov2006}.
    \end{minipage}
\end{table*}

\section{Observations}

We reanalyze the Gemini-North/IGRINS-2 observations of WASP-33~b first presented by C25. The observations were obtained on UT date 2024 January 7 and consist of 48 AB pairs covering orbital phases \(0.399<\phi<0.481\). For the present analysis, we use only the 29 pre-eclipse AB pairs obtained before secondary-eclipse ingress, corresponding to 1.46~hr of observations over \(0.399<\phi<0.451\), when the planetary dayside emission was visible.

We use the same PLP-reduced spectra \citep{Kaplan2024}, secondary wavelength calibration, order selection, and pixel trimming described in C25. We then applied singular value decomposition (SVD) to each spectral order to remove the dominant telluric, stellar, and instrumental variations \citep{deKok2013, Birkby2013, Line2021, Brogi2023}. For the combined volatile and refractory analysis, we removed six singular vectors (SVs), rather than the four adopted by C25, because this configuration maximized the overall recovery of the molecular and atomic signals in the cross-correlation analysis. The inclusion of weaker refractory species shifts the optimal truncation to six SVs, reflecting the species-dependent sensitivity of the recovered signals to SVD detrending. The resulting residual spectra were used for both the detection and retrieval analyses. We adapted the SYSREM residual-RMS diagnostic of \citet{Deibert2021} to SVD and evaluated the RMS before sigma clipping over the same valid samples for one to ten removed components. The RMS decreased rapidly for the first few components and more gradually from approximately six onward, providing a complementary check on the adopted six-component reduction. As a robustness test, we repeated the retrieval using spectra detrended with four and eight SVs and found that the retrieved parameters remain consistent with the fiducial six-SV results within $1\sigma$.

\begin{figure*}[ht!] \centering
\includegraphics[width=0.99\textwidth]{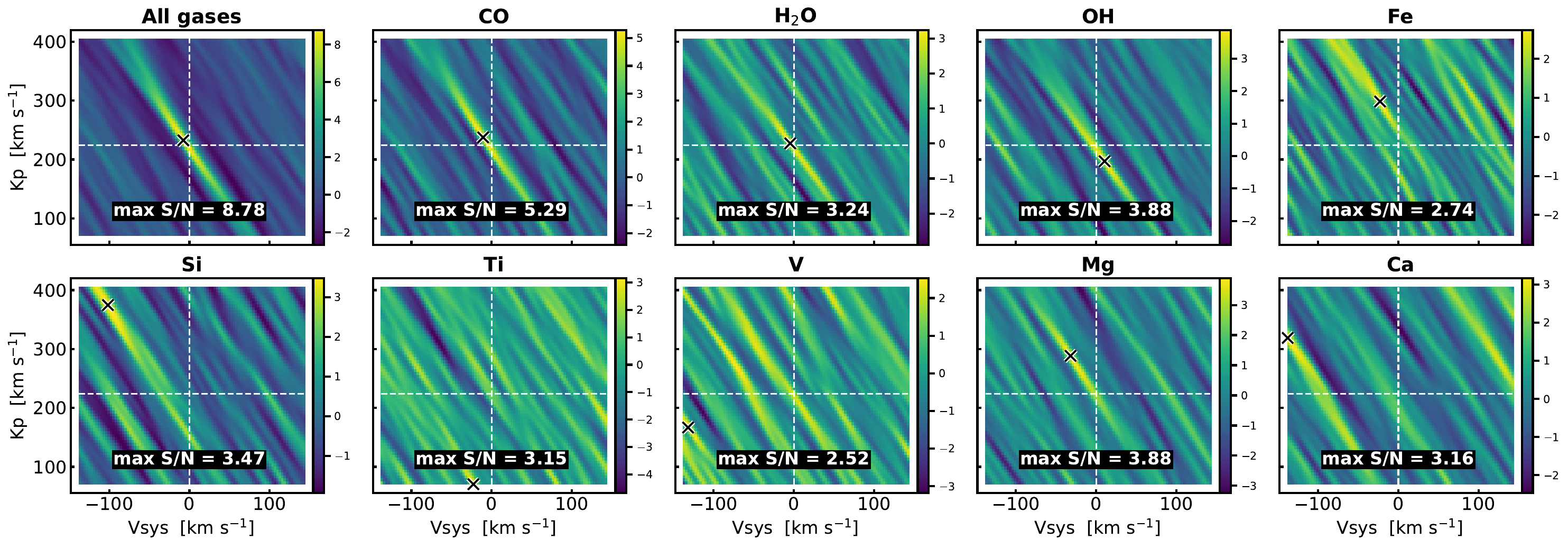}

\caption{
Cross-correlation function (CCF) detection maps using free-chemistry models for all gases and individual species. The maps are shown in the \(K_{\rm p}\)--\(V_{\rm sys}\) space. The horizontal and vertical dashed lines mark the expected planetary velocity and systemic velocity, respectively, and the black crosses mark the maximum-S/N locations in each panel. CO and OH show peaks close to the expected velocities, while H$_2$O shows a weaker nearby feature. \ion{Mg}{1} shows a comparatively strong but velocity-offset peak. Other refractories show no convincing signals at the expected velocities. Each species panel shows the difference between the map computed with the full-opacity model and that computed with the opacity of that species omitted.
}
\label{fig:Free_CCF}
\end{figure*}

\section{Cross-Correlation Analysis} \label{Detection}
The stellar and planetary parameters adopted in this analysis are the same as those used by C25 and are taken from \citet{Chakrabarty2019}. We listed them in Table~\ref{tab:system_parameters}. We utilized \texttt{CHIMERA} \citep{Line2013, Line2021} to generate thermal emission spectra of the dayside atmosphere of WASP-33~b at a spectral resolution of $R=250,000$.
For the detection analysis, we adopted a free-chemistry model in which volume mixing ratios are constant with pressure. The abundances were determined from a grid search initialized around the solar-composition chemical-equilibrium predictions from \texttt{FastChem} \citep{Fastchem1, Fastchem2, Fastchem3}, allowing the abundance of each species to vary independently. The adopted volume mixing ratios are listed in Table~\ref{tab:template_vmr}. For the atmospheric vertical temperature profile, we applied the parameterization from \citet[][hereafter M--S profile]{MS}, using the best-fit temperature and pressure parameters from the retrieval described in Section~\ref{Retrieval}. The opacities\footnote{Cross sections for CO and OH were precomputed using HELIOS-K \citep{HELIOSK}. The H$_{2}$O cross sections were generated as described in \citet{EXOPLINES}. Atomic species were generated using a custom script that accounts for natural, pressure, and Doppler broadening.} chosen for the emission spectra were categorized into three types: (1) continuum sources such as H$_{2}$-H$_{2}$, H$_{2}$-He collision-induced absorption \citep{CIA}, H$^{-}$ bound-free, and H-e$^{-}$ free-free continuum \citep{H-}, (2) volatile molecules such as H$_{2}$O \citep[POKAZATEL;][]{POKAZATEL}, CO, and OH \citep[HITEMP;][]{HITEMP1,HITEMP2}, (3) refractory metals such as \ion{Fe}{1}, \ion{Si}{1}, \ion{Ti}{1}, \ion{V}{1}, \ion{Mg}{1}, and \ion{Ca}{1} \citep{Kurucz}. The model spectra were first convolved with a rotational broadening kernel assuming tidal locking of the system and an equatorial rotational velocity of 6.64\,km\,s$^{-1}$. This was followed by convolution with a Gaussian instrumental profile corresponding to the average resolving power of IGRINS-2 (R = 45,000). The emission spectra for the composite opacities and the opacities for individual species are shown in Figure~\ref{fig:opacities}.

To detect atmospheric signals from candidate species in the dayside atmosphere of WASP-33~b, we performed a cross-correlation analysis \citep{Snellen2010}. Our approach followed previous works using IGRINS and IGRINS-2 \citep[e.g.,][]{Line2021, Brogi2023, Kanumalla2024, Smith2024b, Mansfield2024b, Choi2025}.

\begin{figure*}[ht!] \centering
\includegraphics[width=0.99\textwidth]{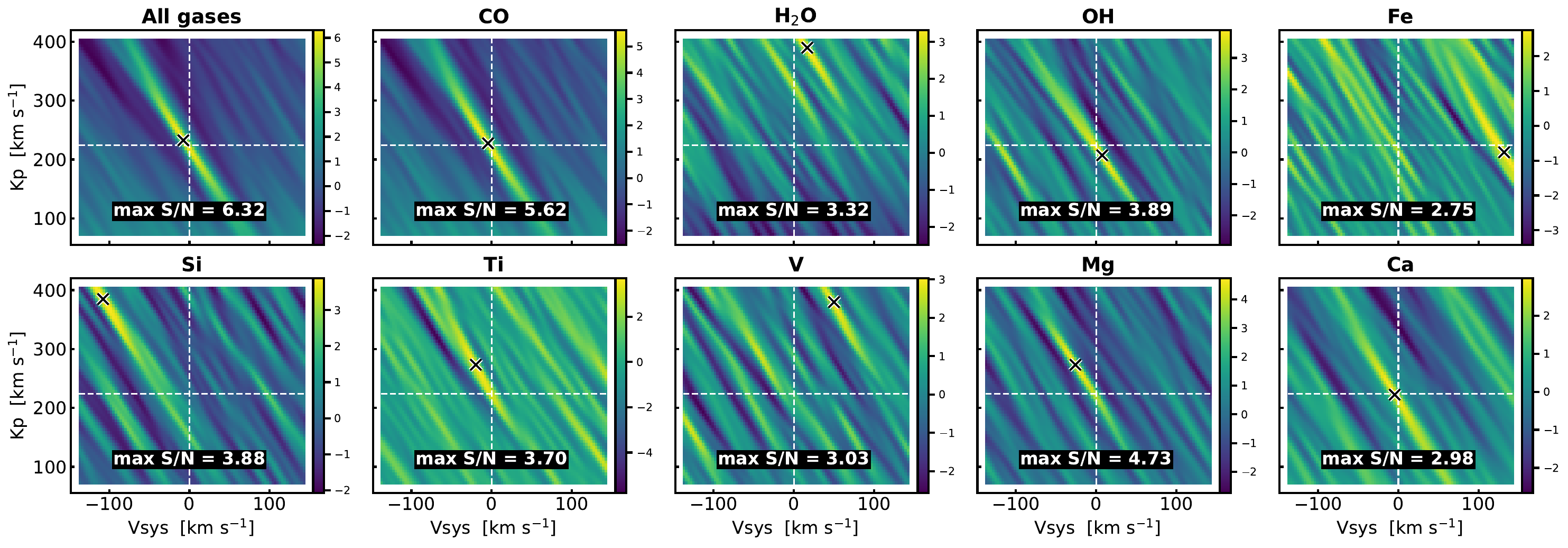}

\caption{
Log-likelihood detection maps from free-chemistry models for all gases and individual species. The plotting convention is the same as in Figure~\ref{fig:Free_CCF}. 
CO is robustly recovered near the expected planetary velocity. \ion{Mg}{1} shows a strong but velocity-offset peak, similar to that seen in the CCF map, while OH shows a weaker feature at a consistent location in both maps. \ion{Ca}{1} shows a tentative peak near the expected velocities, whereas the tentative \ion{Ti}{1} peak is offset from the expected location. H$_2$O, \ion{Fe}{1}, \ion{Si}{1}, and \ion{V}{1} are not detected at the expected velocities. Each species panel shows the difference between the map computed with the full-opacity model and that computed with the opacity of that species omitted.
}
\label{fig:Free_logL}
\end{figure*}

Our detection approach used two methods: traditional cross-correlation with the Pearson correlation coefficient and the likelihood function introduced by \citet{BrogiLine2019}. These were computed across a two-dimensional grid of expected values for the planetary Keplerian velocity semi-amplitude (K$_{p}$) and systemic radial velocity (V$_{sys}$). For each combination of these parameters, the model spectrum was Doppler-shifted accordingly and compared to the data residual matrices. This procedure was applied independently across all usable spectral orders, after which the resulting maps were co-added to construct a combined detection map in velocity space.

These templates are used only for the detection analysis; elemental abundances and the pressure-dependent chemical composition are inferred separately using the chemical-equilibrium retrieval described in Section~\ref{Retrieval}.

The resulting maps from both detection metrics are shown in Figures~\ref{fig:Free_CCF} and \ref{fig:Free_logL}. Figure~\ref{fig:Free_CCF} shows 2D cross-correlation function (CCF) maps, and Figure~\ref{fig:Free_logL} shows a 2D log-likelihood (logL) maps. S/Ns in both maps are computed using the 3$\sigma$-clipping approach introduced by \citet{Kasper2021}. The logL S/N therefore represents an empirical normalization of the logL values. As an additional likelihood-based diagnostic, we calculated velocity confidence intervals following the likelihood-ratio approach of \citet{Brogi2023} and \citet{Panwar2024}. These confidence maps are shown in Figure~\ref{appendix:logL_CI} in Appendix A. To isolate the contributions of individual species to the combined signal, we adopted the approach described in \citet{Smith2024b}: for each gas, we generated a reference model spectrum that includes all opacities except that of the species of interest and recomputed the CCs and logLs. The difference between this and the full-opacity model provides an estimate of the specific contribution each gas makes to the overall signal.

The CCF and logL maps provide complementary diagnostics of the planetary signal. The cross-correlation primarily measures agreement between the patterns of relative line strengths and positions in the data and the template, whereas the likelihood also depends on line contrast, model variance, and detailed line shape. Weak species can therefore appear differently in the two metrics. We consider species recovered when a coherent feature is present in both analyses. H$_2$O is classified as marginal because it is weakly recovered at the expected velocity only in the CC map, whereas the \ion{Ti}{1} and \ion{Ca}{1} features are classified as tentative because they appear only in the logL analysis and lack clear CC counterparts.

We detect the presence of \ion{Mg}{1} in the atmosphere of WASP-33~b. We also recover the volatile species (CO and OH) as presented in C25. While the cross-correlation S/Ns for H$_2$O and OH in our analysis are modest compared to those of C25, we find that the peak positions in our cross-correlation and logL maps exhibit smaller velocity offsets relative to the expected planetary rest frame. This contrasts with C25, where velocity offsets of up to $\sim$50 \kms~were reported. Although the water vapor signal is not recovered in the logL map, it remains marginally visible in the CCF analysis. 

We also identify tentative signals of \ion{Ca}{1} and \ion{Ti}{1} in the logL map, with peak S/Ns of 2.98 and 3.70, respectively. The logL peaks for \ion{Mg}{1} and \ion{Ti}{1} are located at (K$_{\rm p}$, V$_{\rm sys})=(273.1,-25.8$) and ($273.1,-19.7$)~km\,s$^{-1}$, respectively; the origin of these velocity offsets cannot be robustly assessed because of the limited pre-eclipse phase coverage. Given that \ion{Ca}{1} and \ion{Ti}{1} opacities are expected to contribute within the pressure layers probed by IGRINS-2, these tentative features warrant further investigation with higher S/N observations or additional orbital phase coverage for this target. 

Notably, we do not recover significant signals for \ion{Fe}{1}, \ion{Si}{1}, or \ion{V}{1} in either the CCF or likelihood-based analyses. These species have previously been detected in high-resolution observations of WASP-33~b at shorter wavelengths \citep{Nugroho2020, Cont2022a, Cont2022b}, indicating that their non-detection here does not necessarily imply their absence from the atmosphere. Synthetic spectra containing these individual opacity sources show weaker and more blended line structures across the IGRINS-2 bandpass, potentially contributing to a poor match to the residual data cube after detrending. While these non-detections could arise from genuinely low atmospheric abundances or high-altitude ionization, these non-detections may reflect the combination of a lack of strong spectral lines in the IGRINS-2 bandpass relative to the optical and insufficient sensitivity to these species from the relatively short duration of these observations. Continued monitoring with extended orbital phase coverage or additional IGRINS-2 observations may be required to place tighter constraints on their presence.

\section{Bayesian Framework for Atmospheric Retrieval} \label{Retrieval}

To advance beyond the line-detection analysis and place statistical constraints on the atmospheric composition and vertical structure of WASP-33~b, we performed atmospheric retrievals within a Bayesian inference framework. We adopted the same likelihood function used in Section~\ref{Detection} and explored the posterior parameter space using the multimodal nested sampling algorithm implemented in \texttt{pymultinest} \citep{Multinest1, Multinest2}, following a high-resolution retrieval methodology similar to previous studies \citep[e.g.,][]{Line2021, Brogi2023, Smith2024b, Simonnin2025, Kanumalla2026}.

While the detection step (Section~\ref{Detection}) tested the contribution of individual opacity sources, the retrieval analysis is designed to infer a chemically consistent atmospheric state. Given that WASP-33~b is a UHJ subject to extreme irradiation and strong thermal dissociation, we do not adopt a free-chemistry retrieval for the final inference. As discussed by \citet{Brogi2023} and used in \citet{Smith2024b}, free-chemistry retrievals that do not account for thermal dissociation in UHJ regimes can suffer from strong degeneracies and yield chemically implausible solutions compared to an equilibrium-chemistry framework \citep[e.g.,][]{Pelletier2026a}.

Instead, we adopted chemical equilibrium computed with \texttt{FastChem}. For a given P--T structure and set of elemental abundances, \texttt{FastChem} returns the mean molecular weight profile and equilibrium volume mixing ratios at each pressure level, including H, H$^{-}$, e$^{-}$, and H$_2$. We then generated forward emission spectra using CHIMERA, utilizing the equilibrium abundances from \texttt{FastChem} as input opacity profiles. Following \citet{Smith2024b}, we parameterized the composition via individual elemental abundances relative to the solar reference composition of \citet{Asplund2021}.

\begin{table*}
    \centering

    \caption{Parameters and priors used in the atmospheric retrievals for WASP-33~b.}
    \begin{tabular}{ccc}
    \hline\hline
        Parameter & Range & Description \\
        \hline
        \textnormal{[X/H]}  & -3 -- 3 & Elemental abundance (relative to solar) \\
       \hline
       \multicolumn{3}{c}{M--S P--T profile} \\
       \hline
       T$_{0}$  & 500 -- 6000 & Reference temperature [K] \\ 
       log P$_{1}$  & -6 -- 2  & Upper-layer transition pressure [bar] \\
       log P$_{2}$  & -6 -- 2  & Middle-layer reference pressure [bar] \\
       log P$_{3}$  & -6 -- 2  & Deep-layer transition pressure [bar] \\
       $\alpha_{1}$  & 0.01 -- 1.00 & Upper-layer shape parameter \\
       $\alpha_{2}$  & 0.01 -- 1.00 & Lower-layer shape parameter \\
       \hline
       \multicolumn{3}{c}{Bezier-spline P--T profile} \\
       \hline
       T$_0$ & 500 -- 6000 & Temperature of the control point at 10$^{2}$ bar [K] \\
       T$_1$ & 500 -- 6000 & Temperature of the control point at 10$^{0}$ bar [K] \\
       T$_2$ & 500 -- 6000 & Temperature of the control point at 10$^{-1}$ bar [K] \\
       T$_3$ & 500 -- 6000 & Temperature of the control point at 10$^{-2}$ bar [K] \\
       T$_4$ & 500 -- 6000 & Temperature of the control point at 10$^{-3}$ bar [K] \\
       T$_5$ & 500 -- 6000 & Temperature of the control point at 10$^{-6}$ bar [K] \\
       \hline
       $K_{p}$  & 200 -- 250 & Planetary radial-velocity semi-amplitude [\kms] \\
       $V_{\rm sys}$ & -30 -- 30 & Systemic velocity offset [\kms] \\
    \hline
    \end{tabular}
    \label{tab:priors}
\end{table*}

We retrieved the elemental abundances \textnormal{[C/H]}, \textnormal{[O/H]}, \textnormal{[Fe/H]}, \textnormal{[Mg/H]}, \textnormal{[Ca/H]}, \textnormal{[Si/H]}, \textnormal{[Ti/H]}, and \textnormal{[V/H]}, together with the orbital velocity amplitude $K_{p}$ and the systemic velocity $V_{\rm sys}$ (Table~\ref{tab:priors}). Because our observations cover only pre-eclipse phases, we intentionally adopted broad priors on $K_{p}$ and $V_{\rm sys}$ to marginalize over any velocity uncertainty.

A key modeling choice in emission retrievals is the degree of flexibility permitted in the thermal structure. We therefore performed two retrievals that are identical in chemistry and priors, but differ in the P--T parameterization:
(i) the analytic M--S parameterization, and
(ii) a more flexible Bezier spline P--T profile, which was used in \citet{Smith2024b} and \citet{Panwar2025}, and \citet{Kanumalla2026}. This is defined by six temperature nodes at fixed pressures (10$^{2}$, 10$^{0}$, 10$^{-1}$, 10$^{-2}$, 10$^{-3}$, and 10$^{-6}$ bar) and smoothly interpolated in $\log P$.
For both parameterizations, the P--T profile was evaluated on 80 atmospheric layers evenly spaced in $\log P$ between $10^{2}$ and $10^{-6}$~bar.
This second parameterization was included to assess how the inferred composition depends on the flexibility of the adopted P–T profile.

We ran \texttt{pymultinest} with 500 live points for each retrieval. For the comparison between P--T parameterizations, we summarize the number of likelihood evaluations, $N_{\rm evaluation}$, and the Bayesian evidence, $\ln (ev)$, in Table~\ref{tab:posteriors}. From the posterior samples, we also computed derived quantities, including the C/O ratio, an overall metallicity indicator, \textnormal{[M/H]}, and the refractory-to-volatile ratio, \textnormal{[R/V]}. The definition and calculation of \textnormal{[R/V]} are described in Section~\ref{VtoR}.

\begin{figure}[th!] \centering
\includegraphics[width=0.48\textwidth]{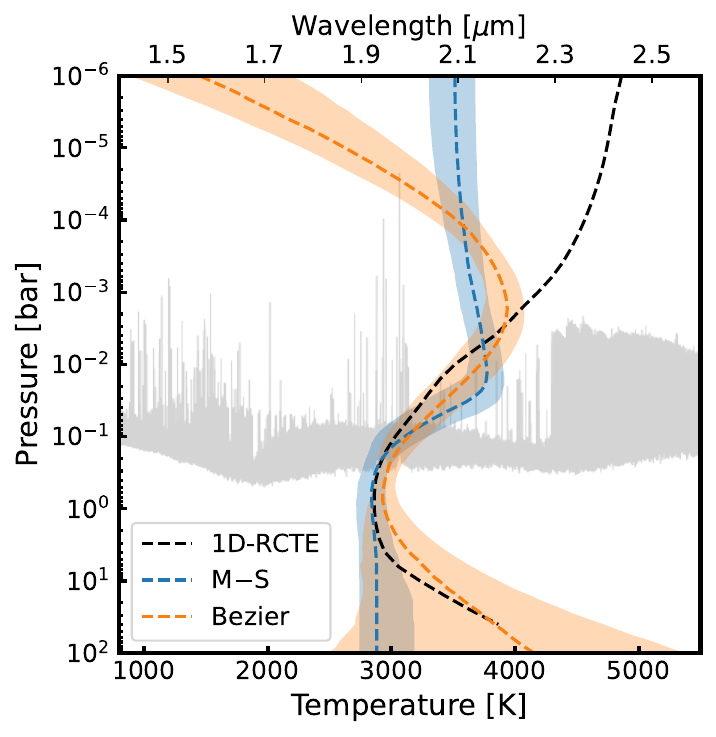}

\caption{
Retrieved P--T profiles for the two retrieval configurations: the analytic M--S parameterization and a six-node Bezier-spline parameterization. Two colored dashed lines show the posterior medians and shaded regions show the 68\% credible intervals. The black dashed line is from the 1D-RCTE profile generated with \texttt{ScCHIMERA}. The light-gray spectrum indicates the \(\tau=2/3\) pressure at each wavelength in the best-fit retrieved atmosphere. 
}
\label{fig:PTs}
\end{figure}

\subsection{Retrieval results}
Here we summarize the constraints on the P--T parameters and elemental abundances obtained from the two retrieval configurations, as listed in Table~\ref{tab:posteriors}. For comparison, we generated a 1D radiative-convective-thermochemical-equilibrium (1D-RCTE) profile with \texttt{ScCHIMERA}, adopting the median C/O and [M/H] values from the preferred Bezier-spline posteriors. 
The retrieved P--T profiles, together with the 1D-RCTE profile, are shown in Figure~\ref{fig:PTs}, and the full corner plot is shown in Figure~\ref{fig:corner} in Appendix~B. 
Both P--T parameterizations recover extremely hot dayside conditions (temperatures exceeding $\sim$3000~K over much of the photospheric region), consistent with expectations for UHJs. 
The Bayesian evidence favors the Bezier-spline parameterization over the M–S profile, with a log-evidence difference of approximately 3.0.

The inferred composition depends strongly on the adopted thermal profile parameterization. 
For the analytic M--S profile, the retrieval favors systematically lower elemental abundances, with \textnormal{[M/H]} $= -0.73^{+0.22}_{-0.16}$, \textnormal{[R/V]} $= -0.41^{+0.29}_{-0.24}$, and ${\rm C/O}=0.68^{+0.07}_{-0.09}$. 
In contrast, the more flexible Bezier-spline retrieval yields higher, less depleted abundances, with \textnormal{[M/H]} $= -0.48^{+0.24}_{-0.18}$, while also giving a higher median \textnormal{[R/V]} $= -0.15^{+0.24}_{-0.23}$ and a consistent ${\rm C/O}=0.65^{+0.09}_{-0.10}$. 
Across the individually retrieved elements, the Bezier profile increases the inferred abundances by $\sim$0.2--0.6~dex relative to the M--S retrieval. 
These differences highlight the sensitivity of the inferred abundances to the adopted P–T parameterization and are consistent with degeneracies between thermal structure and composition.

Among the individually retrieved elements, the most well-constrained elements are volatile elements such as carbon and oxygen ($\sim$ 0.2 dex), and refractory elements such as Mg and Ti ($\sim$ 0.3--0.4 dex) because they show the most significant signals among the refractories in logL maps. Fe and Ca are also well constrained ($\sim$ 0.3 dex) despite their weaker or tentative detection significance in logL maps. This convergence is likely driven by the retrieval being anchored by weak but consistent signals positioned within $K_{p}$--$V_{\rm sys}$ space. Although V is not detected in the cross-correlation analysis, the retrieval yields a comparatively broad [V/H] posterior under the velocity solution shared by all species. The retrieval nevertheless yields abundance constraints despite the absence of clear cross-correlation detections. In contrast, high abundances of \ion{Si}{1} remain disfavored across both thermal parameterizations, consistent with its non-detection in the cross-correlation maps. Consequently, we can only place 3$\sigma$ upper limits on its abundance, yielding \textnormal{[Si/H]} $< 0.81$ and $< 0.24$ for the M--S and Bezier profiles, respectively.

Given its higher Bayesian evidence, we adopt the Bezier-spline retrieval as our preferred model for the subsequent discussion.

\begin{table}

    \footnotesize
    \caption{Retrieved atmospheric parameters from the two Bayesian retrieval configurations.}

    \centering
    \begin{tabular}{ccc}
    \hline\hline
        Parameter & M--S profile & Bezier spline  \\
        \hline
        \textnormal{[C/H]}  & -0.65 $\pm$ 0.19 & -0.42 $\pm$ 0.20  \\
        \textnormal{[O/H]}  & -0.75 $\pm$ 0.19 & -0.49 $\pm$ 0.22  \\
        \textnormal{[Fe/H]} & -1.40 $\pm$ 0.33 & -1.01 $\pm$ 0.31  \\
        \textnormal{[Si/H]} & $<$ 0.81 (3$\sigma$) & $<$ 0.24 (3$\sigma$)   \\
        \textnormal{[Ti/H]} & -0.69 $\pm$ 0.29 & -0.45 $\pm$ 0.31  \\
        \textnormal{[V/H]}  & -0.66 $\pm$ 0.44 & -0.33 $\pm$ 0.54  \\
        \textnormal{[Mg/H]} & -1.21 $\pm$ 0.44 & -0.65 $\pm$ 0.38  \\
        \textnormal{[Ca/H]} & -0.60 $\pm$ 0.30 & -0.21 $\pm$ 0.31  \\
        C/O                 & 0.68$^{+0.07}_{-0.09}$ &  0.65$^{+0.09}_{-0.10}$  \\
        \textnormal{[M/H]}    & -0.73$^{+0.22}_{-0.16}$  & -0.48$^{+0.24}_{-0.18}$  \\
        \textnormal{[R/V]}    & -0.41$^{+0.29}_{-0.24}$  & -0.15$^{+0.24}_{-0.23}$  \\

       \hline
       T$_{0}$     & 3209 $\pm$ 468    & -- \\
       log P$_{1}$ & -1.54 $\pm$ 0.22  & -- \\
       log P$_{2}$ & -0.29 $\pm$ 0.52  & -- \\
       log P$_{3}$ & -0.82 $\pm$ 0.39  & -- \\
       $\alpha_{1}$     & 0.39 $\pm$ 0.21   & -- \\
       $\alpha_{2}$     & 0.077 $\pm$ 0.031 & -- \\
       \hline
       T$_0$ & -- & 4474 $\pm$ 1241  \\
       T$_1$ & -- & 2380 $\pm$ 1141  \\
       T$_2$ & -- & 1620 $\pm$ 927   \\
       T$_3$ & -- & 5213 $\pm$ 672   \\
       T$_4$ & -- & 5293 $\pm$ 529  \\
       T$_5$ & -- & 1117 $\pm$ 509  \\
       \hline
       $K_{p}$     & 224.5 $\pm$ 3.0   &  224.0 $\pm$ 3.6  \\
       $V_{\rm sys}$ & -3.4 $\pm$ 1.4  & -3.1 $\pm$ 1.7  \\
       \hline
       N$_{\rm evaluation}$ & 902439       &  245391   \\
       ln(ev)               & 14971435.45 $\pm$ 0.25 & 14971438.43 $\pm$ 0.22  \\

    \hline
    \end{tabular}
    \label{tab:posteriors}
    
\end{table}

\section{Discussion}\label{Discussion}
\subsection{Comparison with Previous Studies}

The atmosphere of WASP-33~b has been investigated using high-resolution dayside spectroscopy, eclipse spectroscopy, and broadband photometry. These studies consistently favor a hot dayside with a thermal inversion, but the inferred C/O ratio and metallicity differ, partly because of differences in wavelength coverage, chemical assumptions, and P--T parameterizations.

\citet{Finnerty2023} analyzed Keck/KPIC dayside emission spectra at $R\sim35{,}000$, using four spectral orders between 2.2 and 2.5~$\micron$ with substantial gaps. They performed a free-chemistry retrieval using an M--S profile constrained to contain a thermal inversion. They retrieved ${\rm C/O}=0.8^{+0.1}_{-0.2}$ and inferred a metallicity of approximately $2$--$15\times$ solar by comparing the retrieved abundances with equilibrium-chemistry grids. However, they estimated an uncertainty of approximately 1 dex on the metallicity and noted that stellar or substellar values could not be confidently excluded.

\citet{vanSluijs2023} analyzed MMT/ARIES spectra spanning 1.37--2.56~$\micron$ at a measured resolving power of $R\sim15,000$. Rather than performing a continuous elemental-abundance retrieval, they compared the data with grids of self-consistent PHOENIX models, modified PHOENIX models, and spectra derived from a three-dimensional global circulation model. Their preferred model had a metallicity of approximately $10\times$ solar, while subsolar models were disfavored within the explored grid. CO was the only molecule robustly recovered, so the metallicity constraint was primarily driven by the CO line contrast and the available model grid.

\citet{Cont2022b} combined broad optical and near-infrared coverage from CARMENES, HARPS-N, and ESPaDOnS, spanning approximately 0.37--1.71~$\micron$ at resolving powers of $R\sim68{,}000$--$115{,}000$. Their equilibrium-chemistry retrieval was driven primarily by the emission lines of Ti, V, Fe, Si, and OH, with all elemental abundances scaled through a common metallicity parameter. They obtained $[{\rm M/H}]=1.49^{+0.82}_{-0.76}$ dex. The P--T profile was represented by two free pressure--temperature points with isothermal regions above and below them, making it less flexible than the six-node Bezier-spline profile used here. They also reported a strong degeneracy between metallicity and the P--T parameters.

More recently, \citet{Zou2026} combined two CFHT/WIRCam narrowband eclipse measurements with archival HST/WFC3 and Spitzer data. Their equilibrium-chemistry retrieval used a three-node P--T profile, with the pressure of the uppermost node fixed at $10^{-6}$~bar, and yielded ${\rm C/O}=0.78^{+0.03}_{-0.04}$ and $[{\rm Fe/H}]=1.52^{+0.35}_{-0.52}$. These observations constrain the broad dayside emission spectrum but do not resolve the individual molecular and atomic line forests or directly constrain OH.

The present IGRINS-2 analysis provides complementary constraints from simultaneous H- and K-band spectra at $R\sim45{,}000$. Our forward model includes CO, H$_2$O, and OH together with the neutral refractory species Fe, Si, Ti, V, Mg, and Ca. 
The thermochemical-equilibrium framework calculates pressure-dependent abundances, accounting for the strong vertical dissociation and ionization expected in a UHJ atmosphere. We also compare the M--S parameterization with a flexible six-node Bezier-spline profile, allowing the sensitivity of the inferred composition to the assumed thermal structure to be assessed. This provides a more comprehensive joint treatment of the volatile and neutral-refractory inventory than the previous studies considered here, although not every included species is independently detected at high significance.

Our preferred retrieval gives ${\rm C/O}=0.65^{+0.09}_{-0.10}$, which overlaps the lower end of the constraint from \citet{Finnerty2023} but is below the nominal values reported by \citet{Zou2026}. Both the M--S and Bezier-spline retrievals favor subsolar metallicity, although the absolute abundance depends on the adopted P--T parameterization. The preferred Bezier-spline retrieval gives $[{\rm M/H}]=-0.48^{+0.24}_{-0.18}$. Its 95\% credible interval, $[{\rm M/H}]=-0.48^{+0.77}_{-0.35}$, includes the solar value, so a solar metallicity cannot be excluded.

To investigate the metallicity discrepancy between our results and \citet{Cont2022b}, we performed additional equilibrium-chemistry retrievals adopting their two-point P--T parameterization and a common-metallicity approach. We compared the H band alone with the combined H+K spectrum, using identical reduction, likelihood prescriptions, priors on shared parameters, and sampling settings. For the two-point retrievals, we adopted temperature prior ranges of 500–6000 K and pressure prior ranges of \(\log_{10}(P/\mathrm{bar})\in[-6,2]\), with \([\mathrm{M/H}]\in[-3,3]\) and \(\mathrm{C/O}\in[0.01,1.0]\). In the free-scale runs, the dimensionless multiplicative spectral scale \(a\) was fitted with \(a\in[-3,3]\). With the spectral scale fixed at unity, the H-only retrieval yielded $[{\rm M/H}]=1.41^{+1.10}_{-1.11}$, consistent with the enrichment reported by \citet{Cont2022b}, whereas the H+K retrieval yielded $[{\rm M/H}]=-0.21^{+0.32}_{-0.29}$. Allowing the scale to vary shifted the H-only metallicity to $0.53^{+1.09}_{-0.59}$ but left the H+K result essentially  unchanged at $-0.23^{+0.32}_{-0.28}$. All uncertainties denote 68\% credible intervals. The shift toward higher metallicity therefore occurs in the H-only retrievals, whose broader posteriors are more sensitive to amplitude freedom. These results demonstrate that wavelength coverage affects the inferred metallicity within this model: including the K band tightens the constraint and favors a lower value. Thus, a two-point profile does not recover the high metallicity reported by \citet{Cont2022b} when applied to the full H+K dataset.

The differences among reported metallicity estimates should therefore be interpreted in the context of the wavelength coverage and modeling assumptions of each analysis. Previous constraints were derived from CO-dominated infrared spectra, optical neutral-metal lines, or unresolved eclipse fluxes, with different treatments of the thermal and abundance structures. Our additional retrievals illustrate this sensitivity: the H-only analysis yields higher but less tightly constrained metallicities, whereas the combined H+K data yield subsolar median metallicities under the M--S, Bezier, and two-point models. Detrending of high-resolution spectra also removes most broadband continuum information, leaving absolute abundance estimates dependent on line contrasts and their degeneracies with the P--T structure. Thus, although the combined H+K retrievals consistently favor subsolar median values, the absolute metallicity of WASP-33~b remains sensitive to spectral coverage and retrieval assumptions.

\begin{figure*}[ht!] \centering
\includegraphics[width=0.98\textwidth]{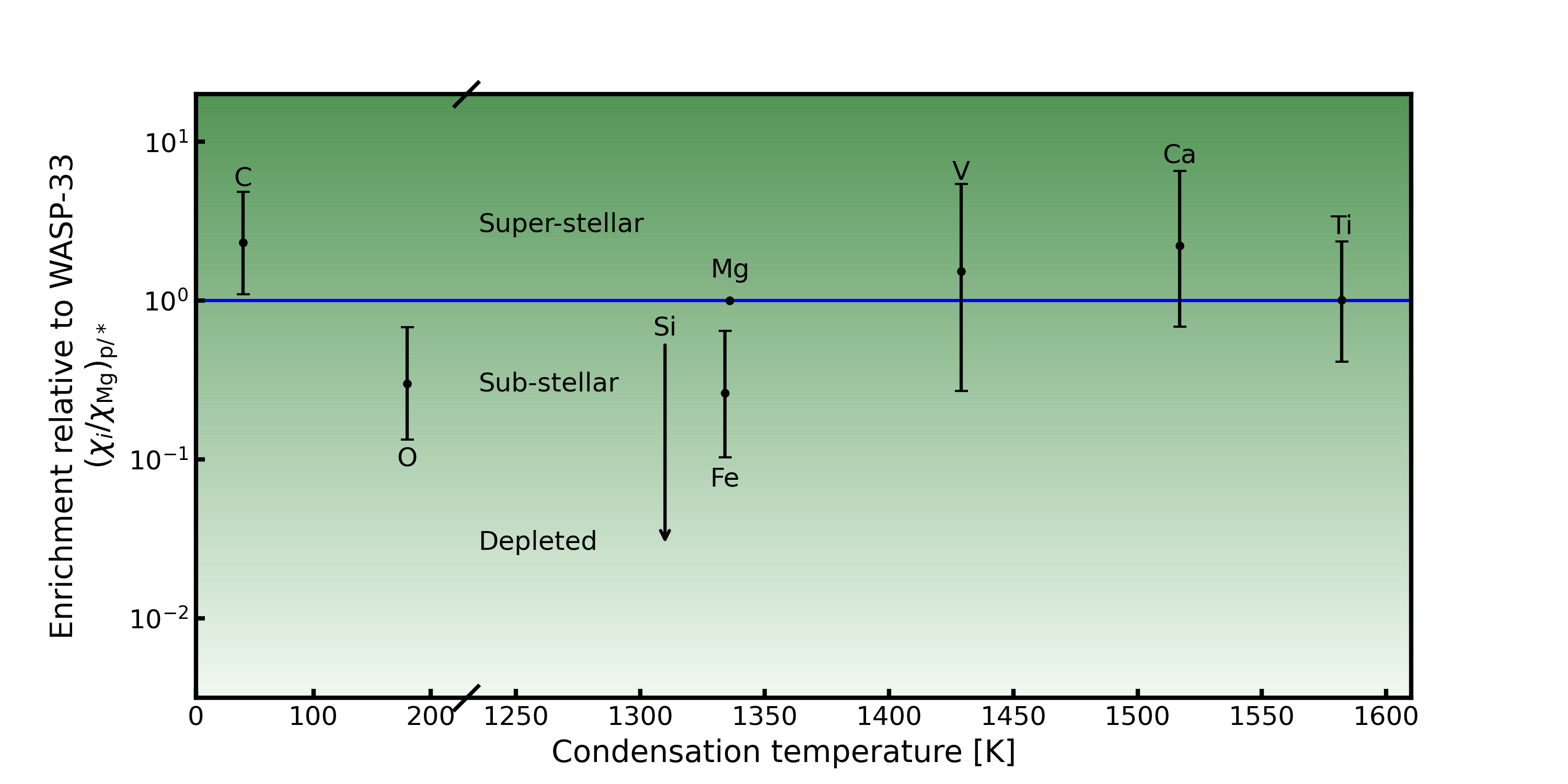}

\caption{
Stellar-normalized elemental abundance ratios relative to Mg, \((\chi_{\rm i} / \chi_{\rm Mg})_{\rm p/ \star}\), as a function of the 50\% condensation temperature. Points and error bars indicate the posterior medians and 1$\sigma$, while the downward arrow shows the 3$\sigma$ upper limit for Si. The horizontal blue line indicates the host-star abundance ratio~\citep{Saffe2021}. We assume a solar V abundance for the host star because no measurement of its [V/H] is available. Condensation temperatures are adopted from \citet{Lodders2003}.
}
\label{fig:XtoMg}
\end{figure*}

\subsection{Abundance Ratios Relative to Mg as Cold-Trapping Diagnostics}

Throughout the following subsections, the subscripts ${\rm p}$ and $\star$ denote the planetary and stellar abundance ratios in solar units, respectively, and ${\rm p}/\star$ indicates normalization to the host-star composition. For the linear ratios in Figure~\ref{fig:XtoMg}, the stellar reference value is unity. In bracket notation, $[X/Y]_{{\rm p}/\star}=[X/Y]_{\rm p}-[X/Y]_\star$, and the stellar reference value is therefore zero. We adopt the stellar abundances from \citet{Saffe2021}.

Refractory-to-refractory abundance ratios provide a useful test of selective refractory depletion. Although absolute metallicity can be model-dependent, relative abundance ratios between elements may be less sensitive to common systematic shifts. Phase-resolved retrievals of KELT-20~b similarly found that the retrieved absolute Fe, Ni, and Ca abundances differed between the pre- and post-eclipse datasets, while the corresponding Ni/Fe and Ca/Fe ratios remained consistent \citep{Bonidie2026}. We use Mg as the reference element rather than the more commonly used Fe, because Fe is not significantly recovered in our detection analysis. Mg also has an abundance and condensation temperature similar to those of Fe, making it a suitable reference element for comparing the retrieved refractory abundance pattern.  Previous works have also shown that the Mg/Fe ratio of ultra-hot Jupiter atmospheres generally reflects the expected stellar value~\citep[e.g.,][]{Pelletier2023, Bazinet2025, Sanchez2026}.

Figure~\ref{fig:XtoMg} shows the stellar-normalized abundance ratios \((\chi_i / \chi_{\rm Mg})_{\rm p/ \star}\) as a function of condensation temperature. Ca and Ti, which have higher condensation temperatures than Mg, do not show the depletion expected from strong cold trapping. In particular, \((\chi_{\rm Ti} / \chi_{\rm Mg})_{\rm p/ \star}\) and \((\chi_{\rm Ca} / \chi_{\rm Mg})_{\rm p/ \star}\) are both consistent with the stellar value within 1 $\sigma$. The non-detected elements by cross-correlation, such as V, Fe, and Si are shown for completeness, but they are not used as primary cold-trapping diagnostics. C and O are also included to provide context for the volatile-to-refractory abundance pattern and are not used as direct diagnostics of refractory cold trapping.

Recent observations suggested that the detection of Ti does not necessarily imply that cold trapping is absent. WASP-76~b shows strong Ti depletion \citep{Tabernero2021, Pelletier2023, Deibert2023}, while Ti has now been detected in WASP-121~b but remains underabundant relative to other refractory species, consistent with partial rather than complete removal by a nightside cold trap \citep{Smith2024b, Prinoth2025, Pelletier2026a}. These results suggest that the relevant diagnostic is the relative Ti abundance, rather than the detection or non-detection of Ti alone.

The population compilation presented by \citet{Pelletier2026b} shows a gradual increase in the observable Ti abundance across the UHJ population: Ti is strongly depleted ($\lesssim$1$\%$) in WASP-76~b, deficient ($\sim$10$\%$) in WASP-121~b, and only mildly underabundant ($\sim$50$\%$) in hotter planets such as WASP-189~b. This onset broadly follows the estimated nightside temperature at which Ti-bearing condensates are expected to vaporize, rather than the equilibrium temperature at which dayside thermal inversions first appear. 
The same study measured the Ti/Fe ratio of WASP-189~b to be between 28\% and 81\% of the stellar value, indicating that partial Ti sequestration may persist even in planets where Ti-bearing species are readily detected.

Our WASP-33~b result is broadly similar to the MASCARA-1~b case, where \citet{Guo2024} retrieved Ti and Fe abundances broadly consistent with solar and found no significant Ti depletion. \citet{Kanumalla2026} also found approximately stellar refractory abundance ratios, providing no strong evidence for nightside cold trapping.
The stellar Ti/Mg ratio measured for WASP-33~b places it toward the hotter end of this observed transition, where significant nightside Ti condensation is no longer expected. 
This interpretation is broadly consistent with the nightside brightness temperatures of $1757\pm88$~K and $1605\pm45$~K inferred from the Spitzer and TESS phase curves, respectively \citep{Zhang2018, vonEssen2020}.
We therefore conclude that the current IGRINS-2 data show no evidence for strong selective Ti cold trapping in the observable pre-eclipse dayside atmosphere of WASP-33~b.  While partial condensation and rainout on the nightside remain possible. Condensates transported back to sufficiently hot dayside layers could re-evaporate and replenish the observable gas phase, unlike the case for colder UHJs such as WASP-76~b and WASP-121~b, where Ti is comparatively depleted relative to other metals.

\subsection{Atmospheric Abundance Ratios for Linking the Formation History} \label{VtoR}

\begin{figure*}[ht!] \centering
\includegraphics[width=0.99\textwidth]{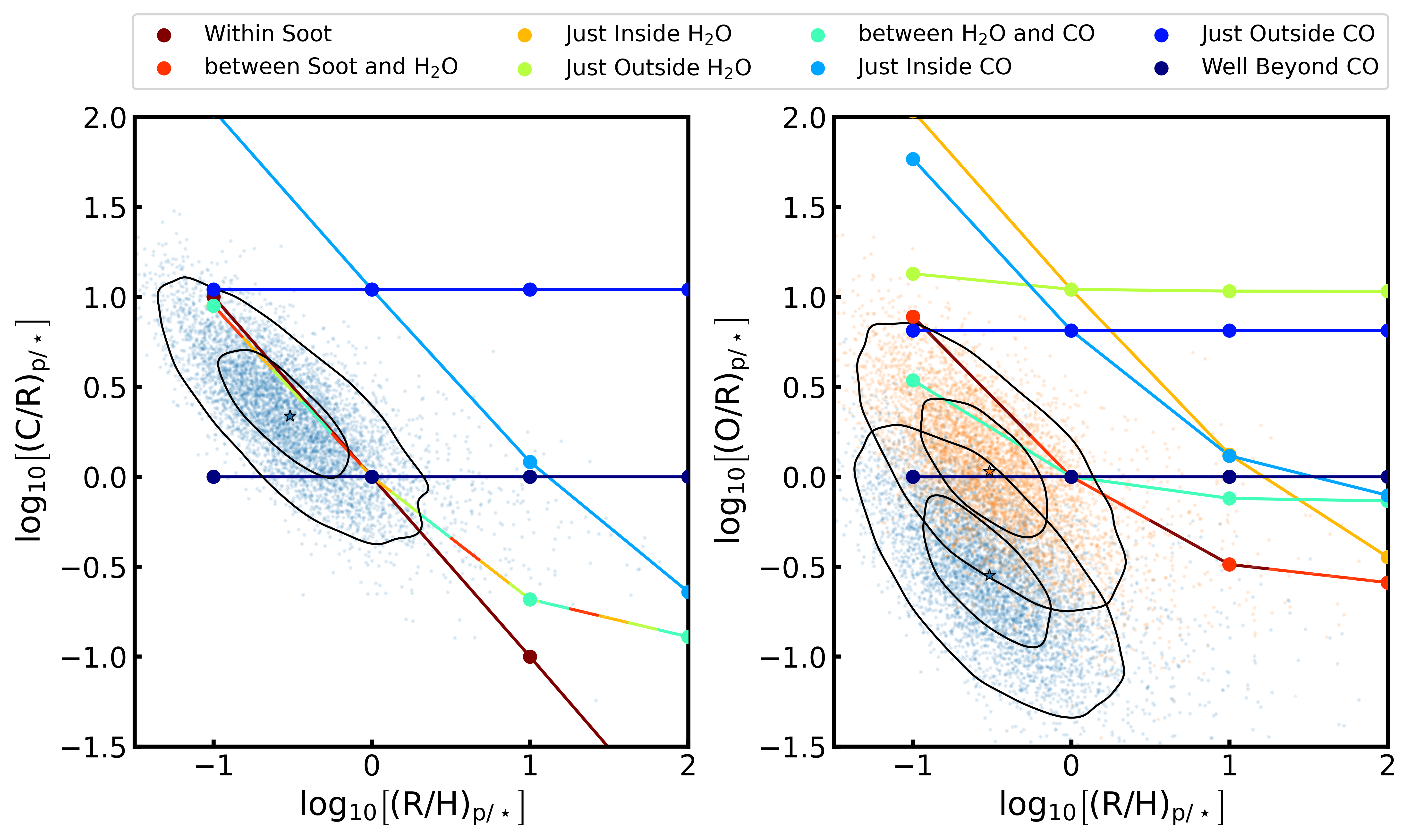}

\caption{
Posterior samples of the stellar-normalized volatile–to–refractory ratios for WASP-33\,b: 
$\log_{10}[(\mathrm{C}/\mathrm{R})_{\rm p}/(\mathrm{C}/\mathrm{R})_\star]$ and 
$\log_{10}[(\mathrm{O}/\mathrm{R})_{\rm p}/(\mathrm{O}/\mathrm{R})_\star]$, 
with $\mathrm{R}=\{\mathrm{Mg,Ca,Ti}\}$. 
Contours denote 68\% and 95\% credible regions. 
Because the stellar oxygen abundance is uncertain, we also show the O/R posterior obtained by normalizing O to the solar abundance while retaining the stellar normalization for the refractory elements (right panel; orange).
Colored lines and markers show the stellar-normalized abundance ratios predicted by the disk-chemistry scenarios of \citet{Chachan2023} for formation at different locations relative to the soot line and the H$_2$O and CO snow lines.}
\label{fig:RtoV}
\end{figure*}

We now compare our inferred volatile-to-refractory ratio for WASP-33~b with the disk-chemistry framework of \citet{Chachan2023}. This comparison is not used to identify a unique formation location, because the current posterior overlaps several formation scenarios within $1\sigma$.

For our retrieved results, we define the volatile inventory as
\[V=N_{\rm C}+N_{\rm O},\]
and the detected refractory inventory as
\[R=N_{\rm Mg}+N_{\rm Ti}+N_{\rm Ca}.\]
We use \(R\), rather than the full refractory inventory, because \ion{Mg}{1} is significantly recovered and \ion{Ca}{1} and \ion{Ti}{1} are tentative, whereas \ion{Fe}{1}, \ion{Si}{1}, and \ion{V}{1} are not significantly detected.

For oxygen, we follow the caution from \citet{Saffe2021} regarding stellar oxygen measurements using the \ion{O}{1} triplet at 7771 \AA, which is prone to NLTE effects. Thus, we treat [O/H]$_{\star}$ values conservatively when interpreting [O/R]$_{\rm {p/\star}}$.

Figure~\ref{fig:RtoV} compares the retrieved stellar-normalized abundance ratios with the qualitative disk-chemistry tracks of \citet{Chachan2023}. The posterior is concentrated at substellar \(R/H\) and generally superstellar \(C/R\), indicating a low detected refractory abundance relative to carbon. However, the 68\% credible region overlaps several formation scenarios, so the current measurements do not identify a unique formation location. The \(O/R\) ratio is additionally affected by the stellar oxygen abundance.

This comparison should therefore be regarded as exploratory. The adopted refractory inventory includes only Mg, Ti, and Ca and does not represent the total refractory content, particularly because Fe, Si, and V are excluded. 
The limited pre-eclipse phase coverage may reduce the sensitivity to weak neutral-metal features, particularly \ion{Fe}{1}, \ion{Si}{1}, and \ion{V}{1}, which have been detected previously \citep{Nugroho2020, Cont2022a, Cont2022b}.

\begin{deluxetable*}{lccccccl}
\tablecaption{Refractory-to-volatile measurements for misaligned UHJs
\label{tab:rv_comparison}}
\tablehead{
\colhead{Planet} &
\colhead{$T_{\rm eq}$} &
\colhead{$g$} &
\colhead{$M_{\rm p}$} &
\colhead{$[R/V]$} &
\colhead{P--T profile} &
\colhead{Reference} &
\colhead{Source} \\
\colhead{} &
\colhead{(K)} &
\colhead{(m\,s$^{-2}$)} &
\colhead{($M_{\rm J}$)} &
\colhead{} &
\colhead{} &
\colhead{composition} &
\colhead{}
}
\startdata
TOI-1431~b  & 2370 & 34.67 & 3.12  & $0.33^{+0.22}_{-0.26}$ & Bezier spline & Solar   & P.~C.~B.~Smith et al., in prep. \\
WASP-121~b  & 2400 & 9.50  & 1.17  & $-0.24\pm0.12$          & Nonparametric & Stellar & \citet{Pelletier2025} \\
WASP-121~b  & 2400 & 9.50  & 1.17  & $0.58^{+0.29}_{-0.25}$ & Bezier spline & Stellar & \citet{Smith2024b} \\
WASP-12~b   & 2590 & 9.67  & 1.47  & $0.13^{+0.29}_{-0.41}$ & Bezier spline & Stellar & \citet{Kye2026} \\
MASCARA-1~b & 2600 & 36.00 & 3.70  & $0.36^{+0.11}_{-0.09}$ & Bezier spline & Solar   & \citet{Kanumalla2026} \\
WASP-189~b  & 2650 & 20.00 & 1.99  & $0.38^{+0.18}_{-0.19}$ & Bezier spline & Stellar & \citet{Sanchez2026} \\
WASP-33~b   & 2780 & 20.00 & 2.093 & $0.37^{+0.38}_{-0.36}$ & Bezier spline & Stellar & This work \\
WASP-33~b   & 2780 & 20.00 & 2.093 & $-0.15^{+0.24}_{-0.23}$ & Bezier spline & Solar   & This work
\enddata
\tablecomments{Uncertainties correspond to the reported 68\% credible intervals. The TOI-1431~b measurement is preliminary.}
\end{deluxetable*}

\subsection{WASP-33~b among misaligned ultra-hot Jupiters with IGRINS and IGRINS-2}
WASP-33~b is strongly spin--orbit misaligned \citep[$\lambda=252^\circ\pm2$,][]{CollierCameron2010}. This orbital architecture is consistent with inward migration after formation away from the present orbit, but it does not identify the migration mechanism. High-eccentricity migration, disk migration followed by later obliquity excitation \citep{Anderson2018b}, secular interactions \citep{Fabrycky2007,Naoz2011}, and primordial disk misalignment \citep{Batygin2012} can all produce spin--orbit misalignment, especially around hot stars where tidal realignment is inefficient \citep{Albrecht2012}. 

\begin{figure}[t] \centering
\includegraphics[width=0.49\textwidth]{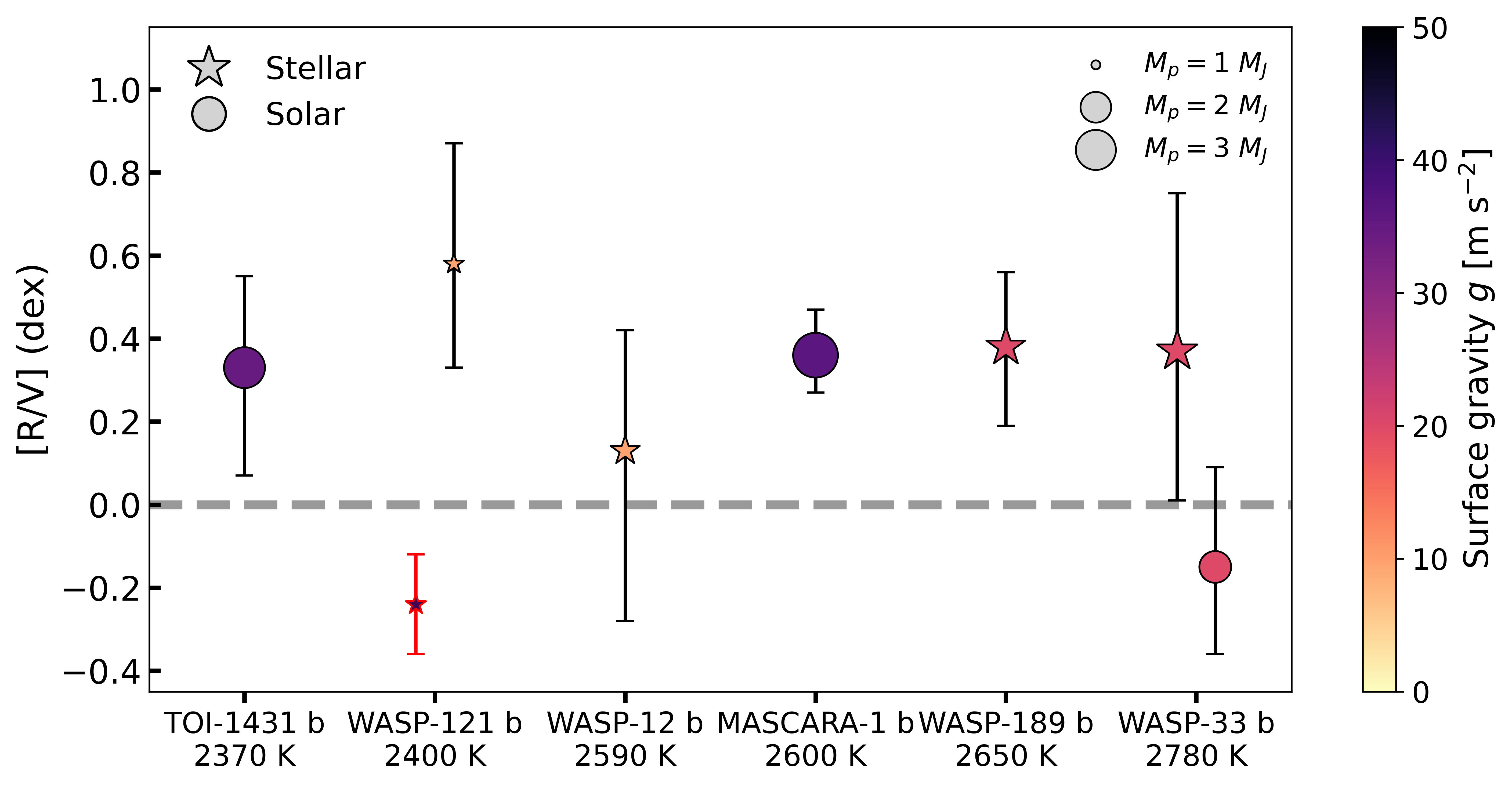}
\caption{
Refractory-to-volatile abundance ratios reported for a sample of misaligned UHJs observed with IGRINS, IGRINS-2, and other instruments. Star-shaped markers indicate ratios normalized to the host-star composition, while circular markers indicate ratios normalized to the solar composition. The horizontal dashed gray line indicates the stellar or solar value. The red marker indicates the value from other instruments \citep[VLT/ESPRESSO and CRIRES$^+$;][]{Pelletier2025}. Marker color indicates the planetary surface gravity, and marker size indicates the planetary mass in Jupiter masses. The equilibrium temperature of each planet is listed below its name. Error bars show the reported 68\% credible intervals. The TOI-1431~b and WASP-12~b values are preliminary results from ongoing analyses and are being prepared for publication. 
}
\label{fig:ReftoVol}
\end{figure}

The refractory-to-volatile ratio, \(R/V\), provides a complementary constraint on the material accreted by the planet. This ratio does not directly determine the migration mechanism, but it can be compared with orbital architecture to test whether chemically inferred formation locations are consistent with the present close-in orbit.
Recent studies have begun to measure refractory-to-volatile ratios for several misaligned UHJs using IGRINS, IGRINS-2, and other high-resolution spectrographs. The reported measurements, together with the planetary properties used in Figure~\ref{fig:ReftoVol}, are summarized in Table~\ref{tab:rv_comparison}. All analyses included in this comparison use flexible P--T parameterizations. Most adopt Bezier-spline profiles, whereas \citet{Pelletier2025} use a nonparametric profile, following \citet{Pelletier2021}. 

Figure~\ref{fig:ReftoVol} compares the reported \(R/V\) measurements across the sample. Most of the IGRINS and IGRINS-2 measurements have superstellar or supersolar median \(R/V\) values. For WASP-33~b, the preferred Bezier-spline retrieval gives $[R/V]_{\rm p}=-0.15^{+0.24}_{-0.23}$ and $[R/V]_{\rm p/\star}=0.37^{+0.38}_{-0.36}$, with the latter being sensitive to the adopted stellar oxygen abundance.

WASP-189~b provides the most direct comparison to WASP-33~b within the current misaligned-UHJ sample. Both planets have surface gravities of \(\sim20~{\rm m~s^{-2}}\), occupy a similar ultra-hot irradiation regime, and have strongly misaligned or near-polar orbits. WASP-189~b has \(\psi\simeq85\)--\(90\) deg \citep{Anderson2018, Lendl2020, Deline2022, Patel2026}, while WASP-33~b has \(\psi\simeq108\)--\(114\) deg \citep{Borsa2021b, Kalman2022, Watanabe2022, Smith2025}. The two systems also show stellar-like refractory-to-refractory ratios in the species that are well constrained: \citet{Sanchez2026} found Mg/Si, Mg/Fe, and Si/Fe to be consistent with stellar values in WASP-189~b, while the Ti/Mg and Ca/Mg ratios of WASP-33~b are consistent with the stellar values. Their \(R/V\) measurements are also consistent within the reported uncertainties.

In this context, WASP-33~b is best treated as an additional comparison point in the emerging UHJ sample rather than as evidence for a universal relation between spin--orbit misalignment and refractory enrichment. A larger and homogeneously analyzed sample is required before any causal relation between obliquity and \(R/V\) can be established.

\section{Conclusion}

We analyzed pre-eclipse dayside emission spectra of WASP-33~b obtained with Gemini-N/IGRINS-2 to investigate its volatile and refractory atmospheric composition. CO produces the strongest signal in both the cross-correlation and likelihood analyses, while \ion{Mg}{1} and OH are recovered at lower significance in both metrics. H$_2$O remains marginal, appearing only in the cross-correlation analysis, and \ion{Ti}{1} and \ion{Ca}{1} show tentative likelihood signals without clear cross-correlation counterparts. \ion{Fe}{1}, \ion{Si}{1}, and \ion{V}{1} are not significantly recovered. The preferred Bezier-spline retrieval gives ${\rm C/O}=0.65^{+0.09}_{-0.10}$ and $[{\rm M/H}]=-0.48^{+0.24}_{-0.18}$. The differences between the Bezier-spline and M--S retrievals demonstrate that the absolute abundances and \(R/V\) ratio are sensitive to the adopted P--T parameterization.

The principal result is the absence of a clear signature of strong selective Ti depletion. The stellar-normalized Ti/Mg ratio is consistent with the host-star value, and Ca/Mg is also consistent with the stellar value within 68\% confidence. WASP-33~b therefore lies toward the hotter end of the observed transition in Ti abundance across the UHJ population, where nightside Ti cold trapping is expected to become less efficient. This result is broadly similar to the approximately stellar refractory ratios reported for MASCARA-1~b. However, the present observations probe only the pre-eclipse dayside and do not exclude partial condensation, rainout, or spatially varying depletion elsewhere in the atmosphere.

The carbon-to-refractory ratios provide only an exploratory constraint on formation. The posterior overlaps several scenarios in the disk-chemistry framework of \citet{Chachan2023}, and the adopted refractory inventory contains Mg, Ti, and Ca rather than the complete refractory budget. We therefore do not infer a unique formation location or migration pathway from the current measurements.

The comparison with other misaligned UHJs provides useful population-level context. Most of the IGRINS and IGRINS-2 measurements have superstellar or supersolar median \(R/V\) values. WASP-33~b extends this pattern to the hottest planet in the current sample, with \([R/V]_{\rm p/\star}=0.37^{+0.38}_{-0.36}\). However, this stellar-normalized value is sensitive to the stellar oxygen abundance; the current measurements do not robustly establish WASP-33~b as refractory-enriched.
A larger sample of aligned and misaligned planets analyzed with consistent refractory definitions, abundance normalization, wavelength coverage, and retrieval assumptions is required before a relation between orbital architecture and refractory enrichment can be established.

\section*{Acknowledgments}
Y.H.C. thanks Woowon Byun for improving the manuscript. 
This research was partly supported by the Korea Astronomy and Space Science Institute under the R\&D program (Project No. 2026-1-904-01, Project No. 2026-1-860-00, Project No. 2026-1-860-01) supervised by the Korea AeroSpace Administration.
This work used the Immersion Grating Infrared Spectrograph 2 (IGRINS-2) developed and built by a collaboration between Korea Astronomy and Space Science Institute (KASI) and the International Gemini Observatory.
Based on observations obtained at the International Gemini Observatory, a program of NSF NOIRLab, which is managed by the Association of Universities for Research in Astronomy (AURA) under a cooperative agreement with the U.S. National Science Foundation on behalf of the Gemini Observatory partnership: the U.S. National Science Foundation (United States), National Research Council (Canada), Agencia Nacional de Investigaci\'{o}n y Desarrollo (Chile), Ministerio de Ciencia, Tecnolog\'{i}a e Innovaci\'{o}n (Argentina), Minist\'{e}rio da Ci\^{e}ncia, Tecnologia, Inova\c{c}\~{o}es e Comunica\c{c}\~{o}es (Brazil), and Korea Astronomy and Space Science Institute (Republic of Korea).
This work was enabled by observations made from the Gemini North telescope, located within the Mauna Kea Science Reserve and adjacent to the summit of Mauna Kea. We are grateful for the privilege of observing the Universe from a place that is unique in both its astronomical quality and its cultural significance.
S.P.\ acknowledges support from the Swiss National Science Foundation under grant 51NF40\_205606 within the framework of the National Centre of Competence in Research PlanetS, the Courtois Scientific Vanguard Fund of the Courtois Foundation, and from the Trottier Institute for Research on Exoplanets. This work was supported by grant https://doi.org/10.69777/377645 from the Fonds de recherche du Québec and by the Center for research in astrophysics of Québec (AstroQuébec).
The National Radio Astronomy Observatory and Green Bank Observatory are facilities of the U.S. National Science Foundation operated under cooperative agreement by Associated Universities, Inc.

\appendix

\section{Velocity confidence intervals map}

\begin{figure}[ht!]
\centering
\includegraphics[width=\textwidth]{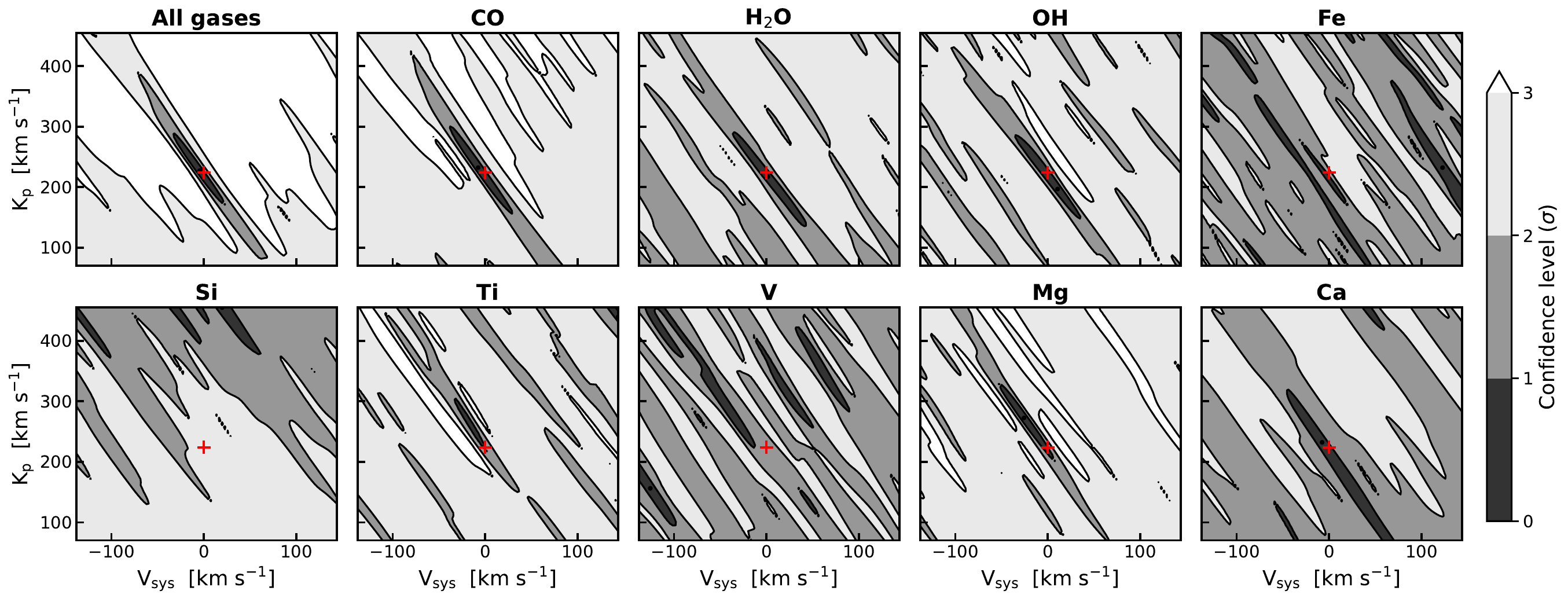}
\caption{
Velocity confidence maps derived from the log-likelihood analysis. The shaded regions and contours show the equivalent Gaussian confidence levels in the K$_{\rm p}$--V$_{\rm sys}$ space. The red crosses mark the expected planetary orbital velocity and systemic velocity.
}
\label{appendix:logL_CI}
\end{figure}

\section{Retrieval Corner Plot}
\label{appendix:corner}

\begin{figure}[ht!]
\centering
\includegraphics[width=\textwidth]{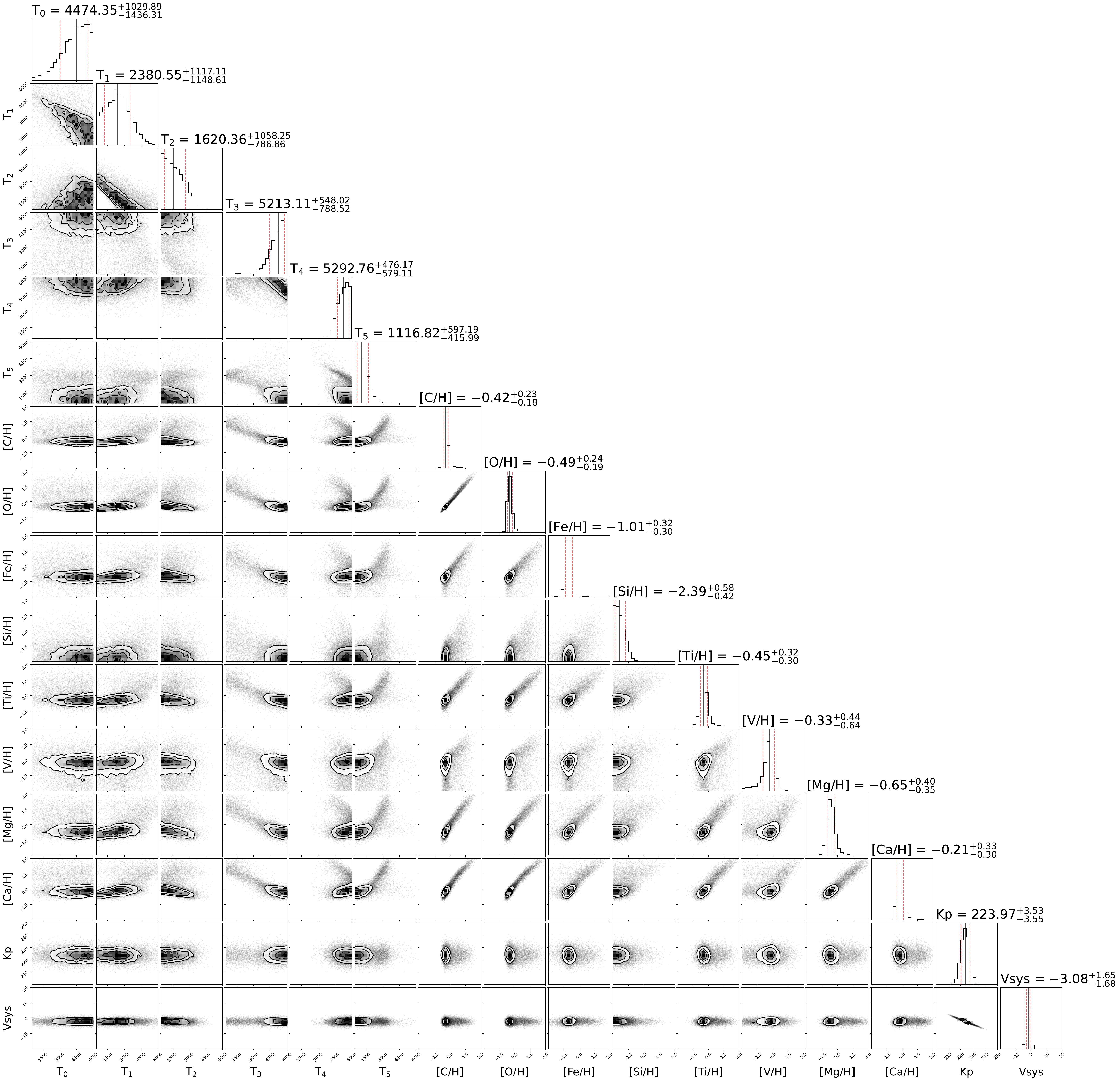}
\caption{
Corner plot showing the marginalized posterior distributions and parameter covariances from the retrieval using the Bezier-spline P--T profile. Retrieved parameters include the pressure-temperature profile, log elemental abundances of C, O, Fe, Mg, Ca, Si, Ti, and V, and velocity parameters ($K_{p}$, $V_{\rm sys}$). 
}
\label{fig:corner}
\end{figure}

\begin{deluxetable}{lc}
\tablecaption{Adopted abundances for the detection templates
\label{tab:template_vmr}}
\tablehead{
\colhead{Species} &
\colhead{$\log_{10}(\mathrm{VMR})$}
}
\startdata
CO          & $-2.7$  \\
H$_2$O      & $-4.6$  \\
OH          & $-6.2$  \\
\ion{Fe}{1} & $-5.0$  \\
\ion{Si}{1} & $-7.7$  \\
\ion{Ti}{1} & $-8.5$  \\
\ion{V}{1}  & $-12.6$ \\
\ion{Mg}{1} & $-5.0$  \\
\ion{Ca}{1} & $-5.5$
\enddata
\end{deluxetable}

\bibliography{sample701}{}
\bibliographystyle{aasjournalv7}


\end{CJK*}

\end{document}